\documentclass[12pt,preprint]{aastex}
\usepackage{placeins}
\usepackage{graphicx}
\usepackage{epstopdf}

\newcommand{\tpol}{T$_{\rm pol}$\ }
\newcommand{\teq}{T$_{\rm eq}$\ }
\newcommand{\rpol}{R$_{\rm pol}$\ }
\newcommand{\rpolnr}{R$_{\rm {pol,nr}}$\ }
\newcommand{\req}{R$_{\rm eq}$\ }

\newcommand{\K}{\,{\rm K}}
\newcommand{\M}{\,{\rm m}}

\newcommand{\Gyr}{\,{\rm Gyr}}
\newcommand{\degs}{\rm degs}

\newcommand{\mas}{\,{\rm mas}}

\newcommand\msun{\hbox{\,M$_\odot$}}
\newcommand\rsun{\hbox{\,R$_\odot$}}
\newcommand\lsun{\hbox{\,L$_\odot$}}
\newcommand\kms{\ \rm km~s$^{-1}$\ }

\newcommand\teff{$\emph{T}_{\mathrm{eff}}$\ }
\newcommand\lapp{$\emph{L}_{\mathrm{app}}$\ }
\newcommand\tapp{$\emph{T}_{\mathrm{app}}^{\mathrm{eff}}$\ }
\newcommand\lbol{$\emph{L}_{\mathrm{bol}}$\ }
\newcommand\tbol{$\emph{T}_{\mathrm{bol}}^{\mathrm{eff}}$\ }
\newcommand\lnr{$\emph{L}_{\mathrm{nr}}$\ }
\newcommand\tnr{$\emph{T}_{\mathrm{nr}}^{\mathrm{eff}}$\ }

\newcommand\fbol{$\emph{F}_{\mathrm{bol}}$}
\newcommand\aproj{$\emph{A}_{\mathrm{proj}}$}

\newcommand\geffbeta{$\emph{g}_{\mathrm{eff}}^{\rm{\beta}}$}
\newcommand\pc{\rm{pc}}
\newcommand\vsini{\rm{$v\sin i$} \ }
\newcommand\wcrit{\rm{$\omega_{\rm crit}$}}
\newcommand\wratio{\rm{$\omega$ / \wcrit}}
\newcommand\lr{$\emph{L-R}_{\rm{pol}}$\ }

\newcommand\betcas{$\beta$~Cas~}
\newcommand\alfleo{$\alpha$~Leo~}

\shorttitle{Rapid Rotators: \betcas \ & \alfleo}
\shortauthors{}

\begin{document}

\title{Colder and Hotter: Interferometric imaging of $\beta$ Cassiopeiae and $\alpha$ Leonis}

\author{X. Che\altaffilmark{1} ,
J. D. Monnier\altaffilmark{1} ,
M Zhao\altaffilmark{2} ,
E. Pedretti\altaffilmark{3} ,
N. Thureau\altaffilmark{3} ,
A. M{\'e}rand\altaffilmark{4,5} ,
T. ten Brummelaar\altaffilmark{4} ,
H. McAlister\altaffilmark{4} ,
S.T. Ridgway\altaffilmark{6} ,
N. Turner\altaffilmark{4} ,
J. Sturmann\altaffilmark{4} ,
L. Sturmann\altaffilmark{4} , }

\altaffiltext{1}{xche@umich.edu: University of Michigan, Astronomy Department,
1034 Dennison Bldg, Ann Arbor, MI 48109-1090, USA}
\altaffiltext{2}{Jet Propulsion Laboratory}
\altaffiltext{3}{University of St. Andrews, Scotland, UK}
\altaffiltext{4}{The CHARA Array, Georgia State University}
\altaffiltext{5}{European Southern Observatory}
\altaffiltext{6}{National Optical Astronomy Observatory, NOAO, Tucson, AZ}

\begin{abstract}

Near-infrared interferometers have recently imaged a number of rapidly
rotating A-type stars, finding levels of gravity darkening
inconsistent with theoretical expectations.  Here, we present new
imaging of both a cooler star \betcas (F2IV) and a hotter one ~\alfleo
(B7V) using the CHARA array and the MIRC instrument at the $\emph{H}$
band.  Adopting a solid-body rotation model with a simple gravity
darkening prescription, we modeled the stellar geometric properties and surface
temperature distributions, confirming both stars are rapidly rotating
and show gravity darkening anomalies.  We estimate the masses and ages
of these rapid rotators on \lr and HR diagrams constructed for
non-rotating stars by tracking their non-rotating equivalents. The
unexpected fast rotation of the evolved sub-giant \betcas offers a
unique test of the stellar core-envelope coupling, revealing quite
efficient coupling over the past $\sim$ 0.5 \Gyr.  Lastly we summarize
all our interferometric determinations of the gravity darkening
coefficient for rapid rotators, finding none match the expectations
from the widely used von Zeipel gravity darkening laws.  Since the
conditions of the von Zeipel law are known to be violated for rapidly
rotating stars, we recommend using the empirically-derived $\beta$ =
0.19 for such stars with radiation-dominated envelopes. Furthermore,
we note that no paradigm exists for self-consistently modeling heavily
gravity-darkened stars that show hot radiative poles with cool
convective equators.

\end{abstract}

\keywords{infrared: stars - stars: fundamental parameters - stars: imaging - stars: individual ($\alpha$ Leonis, $\beta$ Cassiopeiae) - techniques: interferometer}

\section{Introduction}

Stellar rotation is a fundamental property of stars in addition to the
mass and metallicity. However it has been generally overlooked in the
past century for mainly three reasons. Firstly, most stars rotate
slowly. Secondly, no complete stellar rotational model is available to
handle the stellar structure and evolution of a rotating
star. Thirdly, although stellar rotational velocities in the line of
sight \vsini are relatively easy to measure, the inclination angles
are generally unknown, leaving large uncertainties of stellar
rotational velocities.

While almost all cool stars rotate slowly, rapid rotation is the norm
for hot stars. A large fraction of hot stars are observed to be
rotating with equatorial velocities larger than 120 \kms \citep{Abt95,
  Abt02}. Such fast stellar rotation can have strong effects on  the
observed stellar properties. The strong centrifugal forces distort
stellar shapes and make them oblate. Stellar surface temperatures vary
across latitudes due to the gravity darkening \citep{von24a,
  von24b}. Lower effective gravity at the equator result in lower temperatures
compared to the poles. This temperature distribution implies that
apparent luminosities \lapp and apparent effective temperatures \tapp
depend on inclination angles, and the overall values are hidden from
observers. Stellar rotation can also affect the distribution of
chemical elements, mass loss rate and stellar evolution
\citep{mey00}. Some kind of rapidly rotating massive stars may end up as
$\gamma$-ray bursts \citep{mac99}.

Stellar rotation has been studied mainly through the Doppler
broadening line profiles in the past, but the obtained information
from these studies is limited due to the lack of spatial knowledge of
stars, such as the inclination angles. An important and reliable way
to extract such information is through long baseline optical/infrared
interferometry, allowing us to study the detailed stellar surface
properties for the first time. Several rapid rotators has been well
studied using this techniques, including Altair, Vega, Achernar, Alderamin, Regulus and
Rasalhague \citep{van01, van06, auf06, pet06, dom03, mon07, zha09}.

These studies have revealed not only the stellar surface geometry but
also the surface temperature distributions, allowing us to test and
constrain stellar models and laws.  For instance, the surface
temperature distributions have confirmed the gravity-darkening law in
general, but deviate in detail from the standard von Zeipel model
(\teff $\propto$ \geffbeta, where $\beta$ = 0.25 for fully radiative
envelopes). Particularly the studies on Altair and Alderamin prefer
non-standard $\beta$ values from the modified von Zeipel model (the
$\beta$-free model in \cite{zha09}). These results imply the gravity
darkening law is probably only an approximation of the surface
temperature distribution, the real physics behind is still to be
uncovered.

In this paper we show our studies of two rapidly rotating stars with
extreme spectral types in contrast to all the A type stars we have
studied: $\beta$ Cassiopeiae and $\alpha$ Leonis, observed with the
Center for High Angular Resolution Astronomy (CHARA) long baseline
optical/IR interferometry array and the Michigan Infra-Red Combiner
(MIRC) beam combiner. $\beta$ Cassiopeiae ($\beta$ Cas, Caph, HR21)
has $\emph{V}$ = 2.27, \citep{mor78}, $\emph{H}$ = 1.584
\citep{cut03}, 1.43  \citep{duc02}, and is located at $\emph{d}$ = 16.8 \pc
~\citep{van07}. Its mass has been estimated as 2.09 \msun \citep[][see the electronic table on VizieR]{hol07} and it has been
classified as F2III-IV \citep{rhe07}, implying it was an A type star
during main sequence and has evolved  -- here we will present updated mass and luminosity estimates 
(see Section 5).  The rotational
velocity has been reported between \vsini = 69 \kms \citep{gle00} and 82
\kms \citep{ber70} in the literature, although recent measurements are
more consistently confined from 69 \kms to 71 \kms \citep{gle00,
  rei06, rac09, sch09} which we prefer to use in this paper. Previous
studies measured its apparent effective temperature range from 6877\K
to 7200\K \citep{gra01, das03, rhe07, rac09} and estimated its radius
from 3.43 \rsun to 3.69 \rsun \citep{ric02, das03, rac09}.

$\alpha$ Leonis (Regulus, HR3982) has $\emph{V}$ = 1.391
\citep{kha09}, $\emph{H}$ = 1.658 \citep{cut03}, 1.57 \citep{duc02}, distance $\emph{d}$ =
24.31 \pc \ \citep{van07}. It is a well-known rapidly rotating star,
classified as a B7V star \citep{joh53} or B8 IVn \citep{gra03}. The
\vsini measurements from the literature spread a large range from $\sim$
250 \kms \citep{sto84} to $\sim$ 350 \kms  \citep{sle63} and we have
adopted here the recent precise value 317 $\pm$ 3 \kms from \cite{mca05}. Regulus
is also a famous triple star system with the companions B and C
forming a binary system at $\sim$ 175'' away from $\alpha$ Leonis A
\citep{mca05}. Recently \cite{gie08} discovered that $\alpha$ Leonis A
is also a spectroscopic binary with a white dwarf company ($\sim$ 0.3
\msun) of the orbital period $\sim$ 40.11 \rm{d}. The primary mass has
been estimated $\sim$ 3.4 \msun \citep{mca05}, however our study here will show
it is much more massive. The diameter of Regulus has been estimated
several times in the past because of its brightness and relatively
large angular size. \cite{mca05} combined the CHARA K-band
interferometric data and a number of constraints from spectroscopy and
revealed that Regulus has the polar radius \rpol = 3.14 $\pm$ 0.06
\rsun and the equatorial radius \req = 4.16 $\pm$ 0.08 \rsun.

In this paper, we describe the observations and data reduction in
Section 2. Then we show the results of $\beta$ Cas and $\alpha$ Leo
from both the standard and modified von Zeipel models in Section 3. In
Section 4, we present aperture synthesis images. In Section 5, the
locations of the two rapid rotators on \lr and HR diagrams are
discussed. We consider the coupling of the stellar core to the outer
envelope and explore gravity darkening from studying these two rapid
rotators in Section 6.  We conclude in Section 7.

\section{Observation And Data Reduction}
The observations were carried out at the Georgia State University (GSU)
Center for High Angular Resolution Astronomy (CHARA)
interferometer array located on Mt. Wilson. The CHARA array includes
six 1-meter telescopes which are arranged in a Y shape configuration:
two telescopes in each branch. It can potentially provide 15 baselines
simultaneously ranging from 34 \M \ to 331 \M, possessing the longest
baselines in optical/IR of any facility. With these baselines, CHARA offers high
angular resolution up to $\sim$0.4 \mas \ at the $\emph{H}$ band and
$\sim0.7$ \mas \ at the $\emph{K}$ band.

The Michigan Infra-Red Combiner (MIRC) is designed to perform true
interferometric imaging. It is an image plane combiner, combining 4
CHARA telescopes simultaneously to provide 6 visibilities, 4 closure
phases and 4 triple amplitudes. Currently MIRC is mainly used in
$\emph{H}$ band which is further dispersed by a pair of prisms into 8
narrow channels. In order to obtain stable measurements of visibility
and closure phase, MIRC utilizes single-mode fibers to spatially
filter out the atmosphere turbulence. The fibers are arranged on a
v-groove array with a non-redundant pattern so that each fringe has a
unique spatial frequency signature. The beams exiting the fibers are
collimated by a microlens array and then focused by a spherical mirror
to interfere with each other.  Since the interference fringes only
form in one dimension which is parallel to the v-groove, they are
compressed and focused by a cylindrical lens in the dimension
perpendicular to the v-groove to go through a slit of a spectrograph.
The data presented here utilized a pair of low spectral resolution
prisms with \rm R $\sim$ 50. Finally the dispersed fringes are
detected by a PICNIC camera \citep{mon04, mon06}. The philosophy of
the control system and software is to acquire the maximum data readout
rates in real time. The details about the software can be found in
\cite{ped09}.

The integration time is limited by the fast changing turbulence, any
turbulence faster than 3.5 \rm ms readout speed of the camera will
cause decoherence of the fringes. In order to calibrate these fringes,
calibrators with known sizes adjacent to the targets are observed each
night. For the acquisition of true visibility, real time flux of the
beam from each telescope is also required. Several independent methods
(Fiber, Chopper and DAQ , \cite{mon08}) are adopted to indirectly
measure the 'real time' flux.  A recent upgrade of MIRC with
Photometric Channels has been achieved to directly and more accurately
measure the flux. Photometric Channels place a beamsplitter right
after the  microlens array to reflect $\sim$ 25\%
of the flux into multi-mode fibers. The beams exiting the MM fibers go
through the same doublet and prisms, and hit a different quadrant of
the same detector. With Photometric Channels MIRC can now measure the
visibilities with uncertainty down to 3\% \citep{che10}.

We observed \betcas on 7 nights in 2007 and 2009, and \alfleo on 5
nights in 2008. The detailed log is presented in Table 1. Figure 1
shows the overall (u,v) baseline coverage of the observation of \betcas and
\alfleo.

\begin{deluxetable}{llcc}
\tabletypesize{\scriptsize}
\tablecaption{Observation logs for \betcas and \alfleo}
\tablewidth{0pt}
\tablehead{
\colhead{Target} & \colhead{Obs. Date} & \colhead{Telescopes} & \colhead{Calibrators}}
\startdata
\betcas & UT 2007Aug07   & S1-E1-W1-W2  & 7 And \\ 
              & UT 2007Aug08   & S1-E1-W1-W2  & $\sigma$ Cyg, 7 And \\ 
             & UT 2007Aug10   & S1-E1-W1-W2  & $\sigma$ Cyg, 37 And \\ 
             & UT 2007Aug13   & S1-E1-W1-W2  & $\sigma$ Cyg, 7 And, Ups And \\ 
             & UT 2009Aug11   & S1-E1-W1-W2  & 7 And, $\gamma$ Tri \\ 
             	& UT 2009Aug12   & S1-E1-W1-W2  & 7 And, $\gamma$ Tri \\ 
             	& UT 2009Oct22   & S2-E1-W1-W2  & 37 And, $\upsilon$ And, $\epsilon$ Cas, $\eta$ Aur \\ 
\hline
\hline
\\
\alfleo  & UT 2008Dec03   & S1-E1-W1-W2  & $\theta$ Leo \\ 
	    & UT 2008Dec04   & S1-E1-W1-W2  & 54 Gem, $\eta$ Leo \\ 
	    & UT 2008Dec05   & S1-E1-W1-W2  & $\theta$ Hya, $\theta$ Leo \\ 
	    & UT 2008Dec06   & S1-E1-W1-W2  & 54 Gem, $\theta$ Hya, $\eta$ Leo \\ 
	    & UT 2008Dec08   & S1-E1-W1-W2  & $\theta$ Leo \\ 

\enddata

\label{obslog}
\end{deluxetable}

\begin{figure}[thb]
\begin{center}
{
\includegraphics[angle=90,width=3.2in,trim=20mm 0mm 30mm 30mm,clip=true] {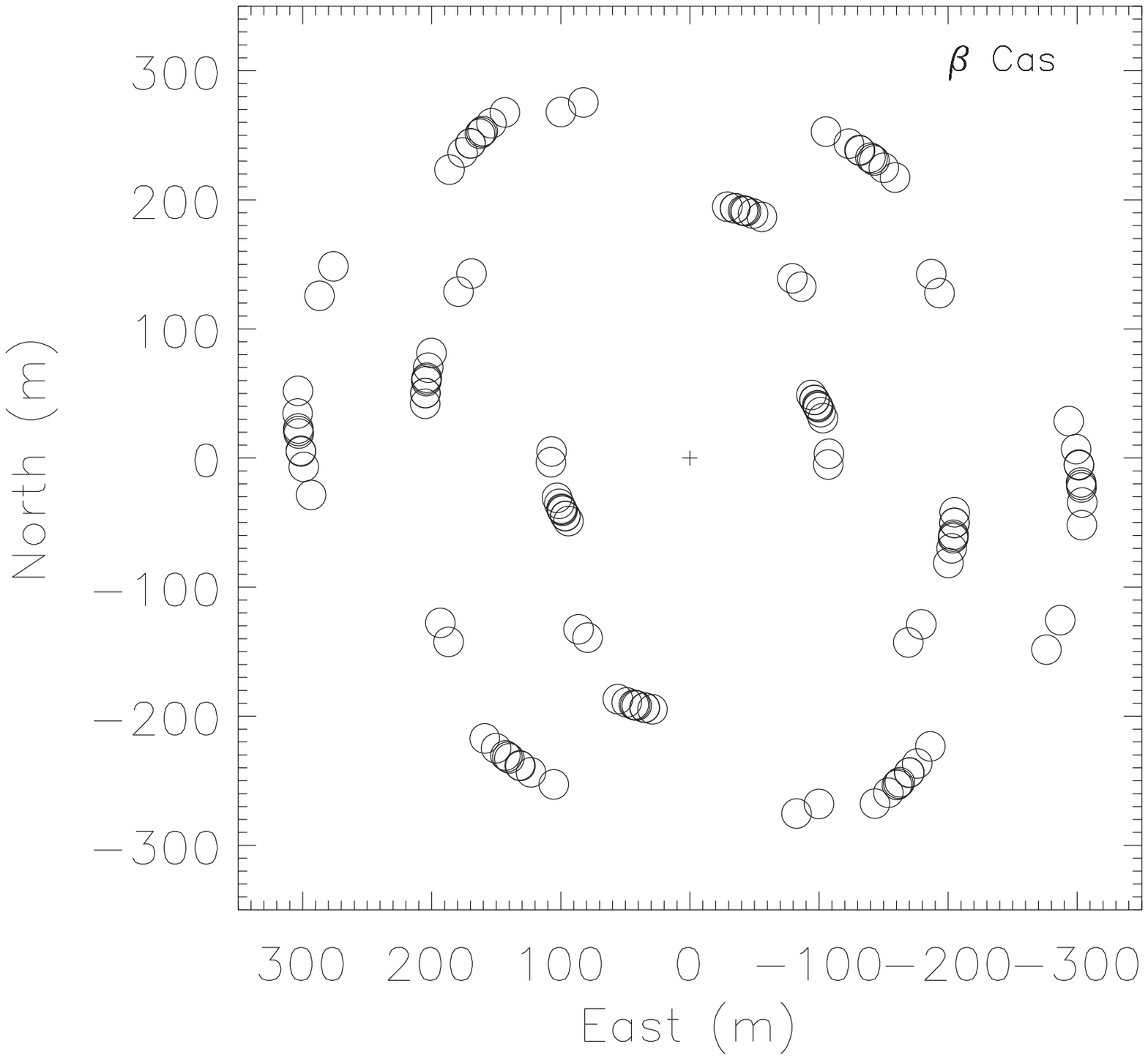}
\includegraphics[angle=90,width=3.2in,trim=20mm 0mm 30mm 30mm,clip=true] {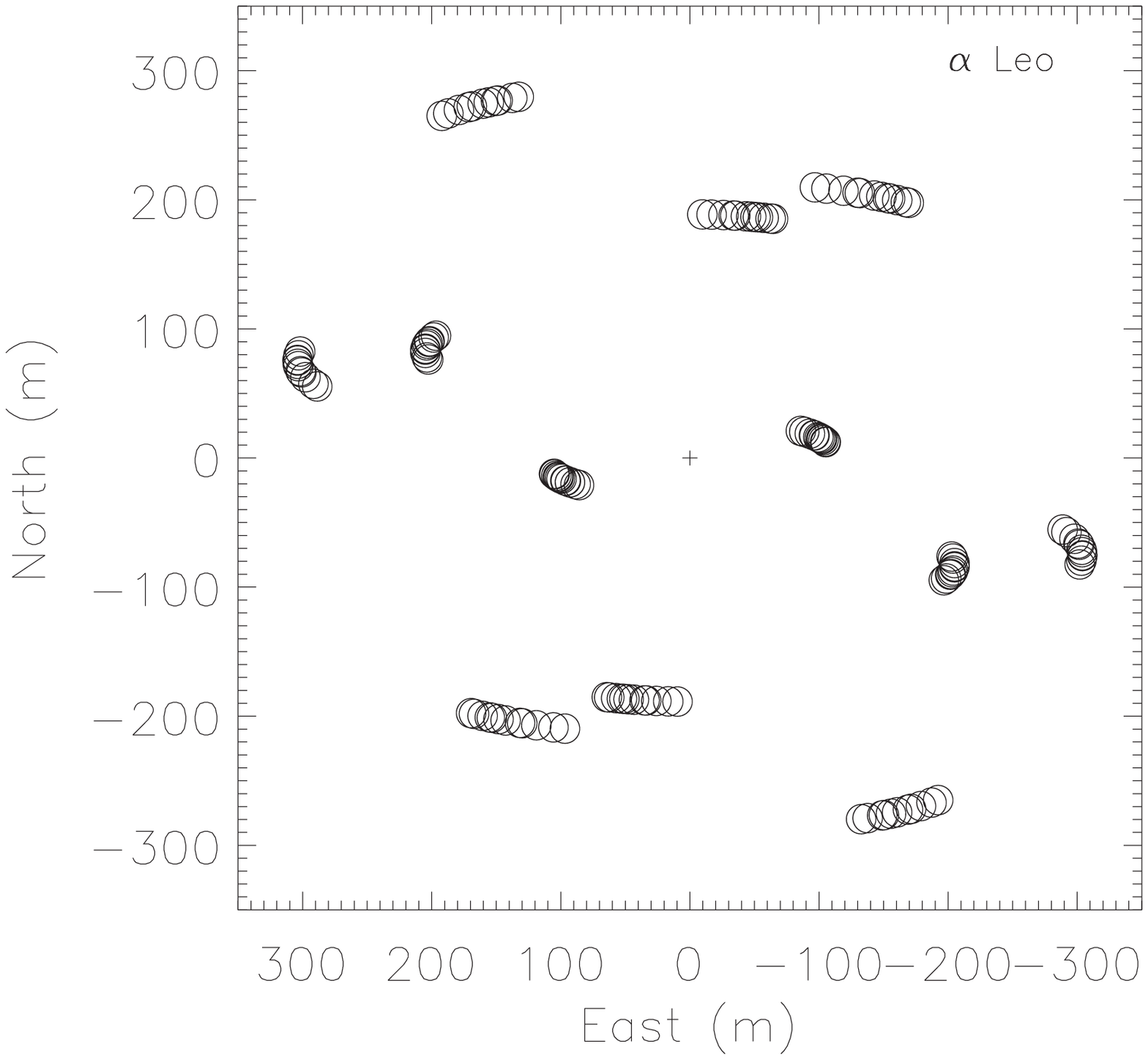}
}
\hphantom{.....}
\caption{ 
Baseline coverage of the all nights observation of \betcas and \alfleo. UV coverage can be calculated by dividing these baselines by the wavelength of $\emph{H}$ band channels.
 \label{uv}}
\end{center}
\end{figure}

Monnier et al. (2007) describes the data reduction pipeline used to
process the data, 
which was validated by using data on the calibration binary
$\iota$ Peg. The pipeline first computes uncalibrated
squared-visibilities and complex triple amplitudes after a series of
background subtractions, Fourier transformations and foreground
subtractions. Then the uncalibrated squared-visibilities and complex
triple amplitudes are calibrated by the fluxes measured simultaneously
with fringes. The calibrators with known sizes are observed to
compensate for the system visibility drift, as listed in Table 2.

\begin{deluxetable}{lcl}
\tabletypesize{\scriptsize}
\tablecaption{Calibrator diameters}
\tablewidth{0pt}
\tablehead{
\colhead{Calibrator} & \colhead{UD diameter (\mas)} & \colhead{Reference} 
}
\startdata
7 And & 0.659 $\pm$ 0.017 & b, c, d\\
37 And & 0.682 $\pm$ 0.030   &  b, c\\
$\upsilon$ And & 1.14 $\pm$ 0.007 & a, b, c, d\\
 $\sigma$ Cyg & 0.542 $\pm$ 0.021 & a\\
 $\gamma$ Tri  & 0.520 $\pm$ 0.0125  & b\\
 $\epsilon$ Cas & 0.351 $\pm$ 0.024  & c, d\\
 $\eta$ Aur & 0.419 $\pm$ 0.063 & c\\
 $\theta$ Leo & 0.678 $\pm$ 0.062 & b, c\\
 $\eta$ Leo & 0.644 $\pm$ 0.068 & c\\
 54 Gem & 0.735 $\pm$ 0.033 & b, c\\
 $\theta$ Hya & 0.463 $\pm$ 0.031 &  c, d\\
 
\enddata

\tablenotetext{a}{\cite{mer08}}
\tablenotetext{b}{\cite{ker08}}
\tablenotetext{c}{\cite{bar78}}
\tablenotetext{d}{\cite{bon06}}

\label{cals}
\end{deluxetable}

\clearpage

\section{Modeling of Rapid Rotators}
We construct a 2D stellar surface model in this paper: the modified
von Zeipel model. The model contains six free parameters,  
stellar polar radius, the polar temperature, the ratio of
angular velocity to critical speed \wratio, the gravity darkening
coefficient ($\beta$), the inclination angle, and the position angle
(east of north) of the pole, to describe the stellar radius, surface
effective gravity and temperature distributions across stellar
surface. The mass of a star is given and fixed in each model fitting
process. Given the stellar mass, stellar polar radius and \wratio, the
stellar radius and surface effective gravity at each latitude can be
determined \citep{auf06}. Then given the stellar polar temperature and
$\beta$, the stellar surface temperature distribution can be computed
from the gravity darkening law(\rm T $\propto$ \geffbeta). Lastly the
orientation of the star is described by the inclination angle and
position angle. In the model, we assume the solid-body rotation for
simplicity; a more complicated and realistic model would consider the
differential rotation which requires additional information (such as
spectral lines) for fitting. The gravity darkening coefficient $\beta$
is a free parameter in the model. By fixing $\beta$, the model reduces
to the standard von Zeipel model ($\beta$ = 0.25, radiative case) or Lucy model
($\beta$ = 0.08, convective case).

In earlier work \citep{mon07}, we found that allow $\beta$ to be a free parameter
greatly improved the fit to the interferometric data.
This flexibility allows us to independently 
test the validity of the standard von Zeipel and Lucy prescriptions. 
Furthermore, the mixture of radiative and convective regions in
the same star may also cause deviations from expected values.
For example, the polar temperature could be thousands of degrees
higher than the equator temperature, resulting in a situation where
upper atmosphere may be radiative at the poles while convective at the
equator. In general, the value of $\beta$ also depends on various
approximations made for the atmosphere, radiation transfer
etc. \citep{cla98}. Therefore in our modified von Zeipel model,
instead of setting $\beta$ to be fixed, we allow $\beta$ to change as
a single free parameter of the model to fit the interferometric
data. For comparison, we also present models with $\beta$ fixed to the
appropriate standard value. The error bars of stellar parameters from
the modified von Zeipel model are in general larger than those from
standard von Zeipel model or Lucy model. This is because there are
certain degrees of degeneracies between the gravity darkening
coefficient $\beta$ and other stellar parameters, as discussed
below.

During the model fitting process, the modified von Zeipel model is
converted into a projected stellar surface brightness model, which is
constrained by the observed $\emph{V}$ and $\emph{H}$
\footnote{We used $\emph{H}$ magnitudes and errors from
only 2MASS catalogue to constrain the model fitting. After we submitted
the paper, we found more precise measurements of H magnitudes \citep{duc02}
which are consistent with our model values within 1-$\sigma$.}
band photometric
fluxes and three kinds of interferometric data from each night:
squared visibilities, closure phases and triple amplitudes. In the
modified von Zeipel model, the stellar surface is divided into small
patches.  The intensity of each patch is computed from Kurucz model
\citep{kur92}\footnote{Data downloaded from kurucz.harvard.edu/}
given
the temperature, gravity, viewing angle and wavelength, so that the
modified von Zeipel model can be converted into the projected
brightness model. The projected brightness model is then converted
into the same three kinds of interferometric data above by direct
Fourier transform to fit to the observed data. We use 4 sub-bands
(binning two adjacent narrow channels dispersed by the MIRC prisms)
across $\emph{H}$ band for accuracy.  In addition, the apparent
$\emph{V}$ and $\emph{H}$ band photometric fluxes are obtained from
the projected brightness model to fit to the observed values. Observed
\vsini is not directly used in the model fitting, but used to
cross-check the results from model fitting. The detailed process is
described in \cite{zha09} and reference therein.

Data errors consist of random errors, errors due to variation of seeing
condition, and
calibration errors from using incorrect diameters of the calibrator targets.
To get the errors from the first two
parts, we treat the data from each night as a whole package and
bootstrap packages randomly with replacement. Then we fit the sampled
data and repeat fifty times to get the distribution of each
model parameter. The upper and lower error bars quoted here are such that the interval
contains 68.3\% probability and the probability above and below the
interval are equal. For the error from the third part, we used simple Monte Carlo sampling
using the our estimated angular size uncertainties -- these errors turned out to be somewhat smaller
than the error from the first two parts.

We should point out that the stellar mass has to be given and fixed at
the beginning of each model fitting process, but at first does not
agree in detail with the model estimated from the fitting results on
both \lr and HR diagrams using the rotational correction (see Section
5). Our approach here has been to adopt the mass from the literature
for the first attempt in the model fitting. The mass estimation from the
first attempt is then used in the second round of model fitting
process. This procedure is repeated until the mass given in the model
agrees with what comes out of the model fitting. The final mass is referred as
the model mass in our paper. The stellar
metallicities are adopted from the literature and fixed
throughout. The distances of the targets are also adopted from the
literature. 

We also calculate the stellar mass based on the measured \vsini range from the literature, 
which is referred as the oblateness mass and was first proposed by  \cite{zha09}. 
For each bootstrap, we extract 
the inclination angle, polar radius and \wratio \ from the best fitting, then uniformly sample
\vsini values 100 times in the given range to obtain a mass distribution. By combining the mass
distribution from each bootstrap, we obtain the whole mass distribution from which the upper 
and lower mass bound can be calculated such that the interval contains 68.3\% probability
and the probability above the upper bound and below the lower bound are the same. 
To compute the best estimation of the stellar mass, we use the best estimations of 
the inclination angle, polar radius and \wratio \ from the model fitting of all nights, and the \vsini value
to be the mean of the measured range from the literature.

\subsection{$\beta$ Cassiopeiae}

We adopted the following  basic properties of \betcas from the literature as
inputs: distance = 16.8 \pc \citep{van07} and metallicity [Fe/H] =
0.03 \citep{gra01}. We take [Fe/H] = 0 which is the closest value to
the observation to extract intensities from Kurucz model. M =
2.09\msun \ \citep[][see the electronic table on VizieR]{hol07} is adopted for the first attempt of the
model fitting. The fitting results and final parameters from the
modified von Zeipel model are shown in Fig. 10 in the Appendix and the
middle column of Table 3 respectively. The results show that \betcas
is rotating more than 90\% of its critical rate, which causes its
radius $\sim$ 24\% longer at the equator than at the poles. The
temperature at the pole is about 1000\K \ higher than that at the
equator. These significant differences between the poles and equator
imply that the \lapp and \tapp are highly dependent on viewing
angles. The best model mass estimation of its non-rotating equivalent
from \lr and HR diagrams is 1.91\msun (Fig. 6), lower than 2.09\msun
from \cite{hol07}. The oblateness mass estimation from \vsini range 69 \kms to 71
\kms is $1.77^{+0.17}_{-0.05}	$\msun, which is consistent with 
our model mass within the error bars. $\beta = 0.146$ from the
modified von Zeipel model fitting is significantly different from
standard values for either radiation-dominated or convection-dominated
envelopes. The inclination angle is low, implying we are looking at
more the polar area than the equatorial area as shown in Figure 4 (see
Section 4). This is why the apparent luminosity \lapp is higher than
\lbol.  

\cite{cla00} has computed the evolution of gravity darkening
coefficients for different stellar masses, and showed that at such low
\teff \ as \betcas it should be convection-dominated in the
envelope. Fixing gravity darkening coefficient $\beta = 0.08$ (Lucy
model) for convective envelopes, we run model fitting again and the
results are shown in the right column of Table 3. The
best fitting  $\chi^2$s for this model is much worse, nearly a factor of 2 higher. 
Many parameters from the Lucy model are similar to those from the modified von Zeipel
model, except the temperature at the equator. This is not surprising
because the low $\beta$ value means the weak dependence of the
temperature on gravity, namely the temperature at the equator will be
closer to that at the poles for the Lucy model. Consequently the
luminosities and temperature \lapp, \lbol and \tapp are a little
higher than those from the modified von Zeipel model. The modified von
Zeipel model gives significantly lower $\chi^2$ than the Lucy model,
especially that from the closure phase data which is sensitive to
asymmetric structures on the stellar surface. This implies the
modified von Zeipel model describes the surface temperature
distribution better, ruling out the Lucy model in this case. This is
also confirmed by comparing the model \vsini with the observed values:
\vsini = $72.4^{+1.5}_{-3.5}$ \kms from the modified von Zeipel model 
agrees with the observation 69 \kms to 71 \kms, 
while from the Lucy model \vsini =  $81.3^{+0.9}_{-1.0}$ \kms deviates strongly
from the observation. Further more, the oblateness mass and model mass 
don't agree with each other, suggesting that the Lucy model is not self-consistent
in this case.

We found that the low inclination angle induces strong degeneracies
between some parameters during the model fitting. For example when a
star is pole-on the darkness at the equator could be due to either the
high angular velocity or the high gravitational darkening
coefficient since the oblateness can not be directly constrained from this viewing angle. 
Therefore we explore the probability spaces of gravity darkening coefficients $\beta$
with inclination angles and \wratio \ to assess possible correlations. For example, we first
search the best model fitting results of all nights on a 40 $\times$ 40 grid of $\beta$ 
and inclinations by fixing these two parameters on each pixel. 
Generally if these two parameters are independent, then the probability of 
their true values falling into each pixel is $\propto$
$e^{-0.5\chi^2}$. However in this case the two parameters are dependent, we modify the
probability $\propto$ $e^{-\alpha\chi^2}$, where $\alpha$ is a variable to be determined.
Then we overplot the results of the two parameters from each bootstrap
onto the probability space (not shown in the figure), and find the contour of the same
$\chi^2$ containing 68.3\% of bootstrap results, from which $\alpha$ can be computed.
The contour is defined as 1-$\sigma$.

The left panel of Fig. 2 shows the degeneracy between $\beta$ and the
inclination. The contour represents the 68.3\% probability level, and
is weakly elongated in one direction. We further overplot onto the
probability space the observed \vsini range which intersects the contour. This
means a precise \vsini measurement would significantly constrain the
stellar parameters from our model fitting. The same idea is applied to
the probability space of $\beta$ and \wratio \ (Figure 2 right) which
shows a stronger correlation between these two parameters.

\begin{figure}[thb]
\begin{center}
{
\includegraphics[width=7in,trim=27mm 20mm 0mm 10mm,clip=true] {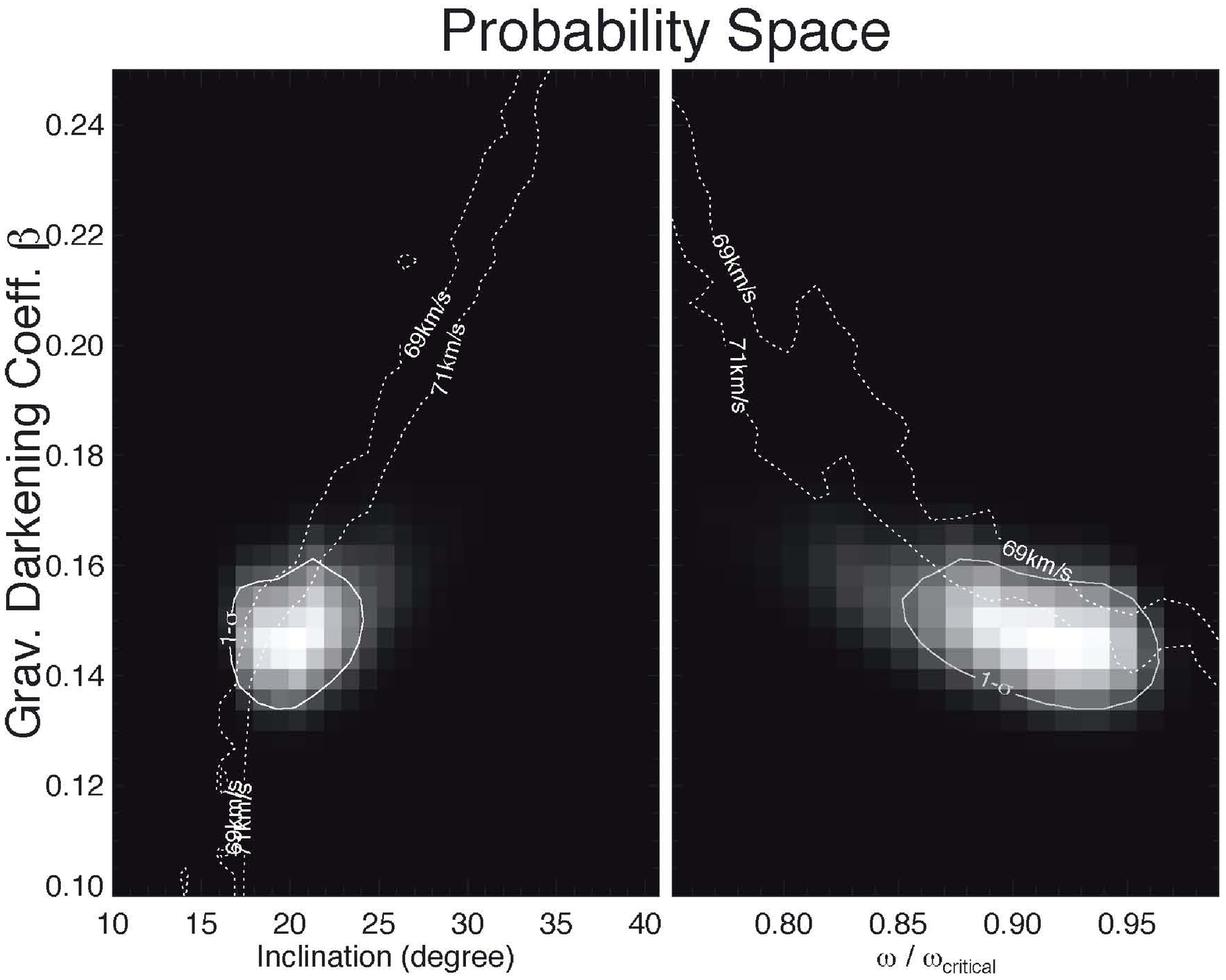}
}
\hphantom{.....}
\caption{ Probability spaces of $\beta$ Cas show the degeneracy
  between stellar parameters. The left panel shows the probability
  space of the gravity darkening coefficient $\beta$ and the
  inclination angle; the right one shows that of $\beta$ and the
  fraction of critical angular velocity \wratio. The solid contours
  represent the 1-$\sigma$ levels, containing 68.3\% of the
  probability. And the dashed lines connect pixels in the probability
  space with the same \vsini values from model fitting. The value
  range 69 \kms to 71 \kms is adopted from the literature, and the
  corresponding lines intersect the 1-$\sigma$ contours. Both panels
  show the elongation of the contours, which imply some degeneracies
  between these parameters.
 \label{betacas1}}
\end{center}
\end{figure}

\begin{deluxetable}{lcc}
\tabletypesize{\scriptsize}
\tablecaption{Best-fit and physical parameters of \betcas}
\tablewidth{0pt}
\tablehead{ 
\colhead{Model Parameters}
&\colhead{Modified von Zeipel model ($\beta$-free)}
&\colhead{Lucy model ($\beta$ = 0.08)}
}
\startdata
Inclination (\degs)         			& $19.9^{+1.9}_{-1.9}$ 					& $21.4^{+3.1}_{-0.9}$				 \\
Position Angle (\degs) 			&  $-7.09^{+2.24}_{-2.40}$				& $-1.8^{+0.8}_{-1.7}$				 \\
\tpol (\K)		                			&  $7208^{+42}_{-24}$ 					& $7108^{+14}_{-18}$ 			\\
\rpol (\mas)	          			&  $0.849^{+0.023}_{-0.020}$          			&$0.835^{+0.035}_{-0.010}$			\\
\wratio    						&  $0.920^{+0.024}_{-0.034}$ 				& $0.930^{0.011}_{-0.050}$	 			\\
$\beta$                          			& $0.146^{+0.013}_{-0.007}$	 			& 0.08 (fixed)				\\

\hline
\hline
Derived Physical Parameters		&							&						\\
\hline
\teq  (\K)                				&  $6167^{+36}_{-21}$ 				& $6487^{+12}_{-17}$ 			\\
\req  (\rsun)    					&  $3.79^{+0.10}_{-0.09}$ 			& $3.77^{+0.16}_{-0.04}$ 			\\
\rpol (\rsun)   					&  $3.06^{+0.08}_{-0.07}$				& $3.01^{+0.13}_{-0.04}$			 \\
Bolometric luminosity \lbol (\lsun) 	& $21.3^{+1.0}_{-0.7}$ 				& $22.7^{+1.4}_{-0.3}$				 \\
Apparent effective temperature \tapp(\K)	& 6825 							& 6897					\\
Apparent luminosity \lapp(\lsun) 	& 27.3	 							& 28.3						\\
Model \vsini (\kms )\tablenotemark{a}	& $72.4^{+1.5}_{-3.5}$ 				& $79.8^{+0.9}_{-1.0}$		 		\\
Rotation rate (rot/day)			&$1.12^{+0.03}_{-0.04}	$			&$1.16^{+0.01}_{-0.06}$			\\
Model mass (\msun)\tablenotemark{b}	&$1.91 \pm 0.02$				&$1.95 \pm 0.03$			\\
Oblateness mass (\msun)	\tablenotemark{c}	& $1.77^{+0.17}_{-0.05}	$			&$1.45^{+0.12}_{-0.27}$		\\
Age (Gyrs)\tablenotemark{b}		&$1.18 \pm 0.05$					&$1.09 \pm 0.03$			\\
Model $\emph{V}$ Magnitude\tablenotemark{d}    & $2.284^{+0.012}_{-0.019}$ 						& $2.251^{+0.020}_{-0.006}$					\\
Model $\emph{H}$ Magnitude\tablenotemark{d}    & $1.398^{+0.007}_{-0.007} 	$					& $1.394^{+0.010}_{-0.001}$					\\

\hline
\hline
$\chi^{2}$ of various data			&							&						\\
\hline
Total $\chi^2_{\nu}$        			&1.36  	 					& 2.53 					 \\
Vis$^2$ $\chi^2_{\nu}$ 			&1.26   						& 1.56 					 \\
CP $\chi^2_{\nu}$          			& 2.18    						& 4.81					\\
T3amp $\chi^2_{\nu}$   			& 0.45    						& 0.60					\\

\hline
\hline
Physical Parameters from the literature	&							&						\\
\hline
[Fe/H]\tablenotemark{e}  & \multicolumn{2}{c}{0.03} \\
Distance (\pc)\tablenotemark{f} & \multicolumn{2}{c}{16.8} \\

\enddata
\tablenotetext{a}{Observed \vsini = 69 \kms to 71\kms \citep{gle00, rei06, rac09, sch09}}
\tablenotetext{b}{Based on the $Y^2$ stellar evolution model \citep{yi01, yi03, dem04}}
\tablenotetext{c}{\cite{zha09}}
\tablenotetext{d}{\emph{V}mag = 2.27 $\pm$ 0.01, \citep[][with arbitrary error]{mor78}, 
\emph{H}mag = 1.584 $\pm$ 0.174 \citep{cut03}, 1.43 $\pm$ 0.05 \citep{duc02} }
\tablenotetext{e}{\cite{gra01} }
\tablenotetext{f}{\cite{van07} }
\label{betcas_tab}
\end{deluxetable}

\subsection{$\alpha$ Leonis}
We first fit the stellar surface of the modified von Zeipel model to
the interferometric data of \alfleo. The parameters we adopted from
the literature are given as following: distance = 24.31 \pc
\ \citep{van07}, metallicity [Fe/H] = 0.0 \citep{gra03}. Mass = 3.4
\msun \citep{mca05} was used for the first attempt of the model
fitting. The fitting results from the modified von Zeipel model are
shown in Fig. 11 in the appendix, with the final stellar parameters
listed in the middle column of Table 4. \alfleo is rotating at 96\% of
its critical speed, causing the equatorial radius about 30\% longer
than the polar radius. The temperatures at the poles are more than
3000K hotter than that at the equator. The gravitational darkening
coefficient $\beta$ from the fitting is again different from the
"standard" values for either radiative or convective envelopes. The
results show that \alfleo is almost equator-on, which is shown as a
dark strip in Figure 5 (see Section 4).  Therefore the \lbol is higher
than the \lapp. The model mass
from HR diagram is 4.15 $\pm$ 0.06 \msun. Adopting the \vsini range
\vsini = 317 $\pm$ 3 \kms 
from \cite{mca05} paper, the oblateness mass estimation corresponding
to the model mass is 
$3.66^{+0.79}_{-0.28} $\msun, which also agrees with the model mass
within the errors. The large errors of the oblateness mass is due to the degeneracy
of stellar parameters as discussed later. 
The observed \vsini \citep{mca05}
is consistent with our derived value $336^{+16}_{-24}$ \kms with error bars.  

Theoretically the high surface temperature of \alfleo suggests that
the envelope is fully radiative, corresponding to the gravity
darkening coefficient $\beta = 0.25$. We fit the model again using the
fixed $\beta$ value, which is the standard von Zeipel model. The best
fitting $\chi^2$s for this model is much worse, nearly a factor of 2
higher.  For completeness, we have included the results in the right
column of Table 4. In this scenario, \alfleo is rotating even
faster. The larger gravitational darkening coefficient and faster
rotation imply even larger temperature difference between the poles
and equator. However the derived equatorial temperatures from the
modified and standard von Zeipel models agree with each other. This is
because Regulus is almost equator-on, the observed interferometric
data is dominated by information from the equator. The $\chi^2$s of
the various interferometric data from the modified von Zeipel model
are all significantly smaller than those from the standard von Zeipel
model, supporting the modified von Zeipel model with $\beta$ = 0.19 is
preferred to describe the surface properties of Regulus, ruling out the 
standard von Zeipel value. This conclusion is also supported by the disagreements 
between the model mass and oblateness mass from the standard von Zeipel model, 
and between the model and observed \vsini values (see Table 4).

We expect some degeneracies of parameters from the modified von Zeipel
model fitting because of the symmetry of the equator-on
orientation. Two figures of probability space of \wratio and the
inclination vs. $\beta$ are shown in Figure 3. Both pictures show a
strongly elongated contour of the probability, implying significant
correlation between these parameters. The solid contours show the
68.3\% probability. We overplot the observed \vsini range from
\cite{mca05}, which intersects the contour with a much smaller common
area. Therefore a precise \vsini measurement would significantly
reduce the degeneracy between the parameters and constrain them much
better.

Based on only visibility data, \cite{mca05} modeled \alfleo and our
new model results are generally consistent with this earlier
work. Since MIRC has higher angular resolution, better UV coverage and
the closure phase data, our data is more sensitive to the detailed
structures such as the inclination and position angles. This work
found acceptable fits for $\beta$ values between 0.12 and 0.34 (best
fit at 0.25), a range consistent with our more refined analysis.
Although our estimations of the bolometric luminosity \lbol of Regulus
is similar to those from their paper, the HR diagram (Fig. 7) from our
results suggests that the mass of the non-rotating equivalent of
Regulus is 4.15 $\pm$ 0.06\msun, much more massive then the 3.4 $\pm$
0.2\msun that \citet{mca05} obtained using the surface gravity $log~g$
from spectral analysis.  Their results show that the non-rotating
equivalent of Regulus has lower mass and consequently lower \lbol than
rapidly rotating Regulus, which is in contrast to what \cite{sac70}
found that a non-rotating equivalent actually has higher \lbol than
its rapidly rotating equivalent.

\begin{figure}[thb]
\begin{center}
{
\includegraphics[width=7in] {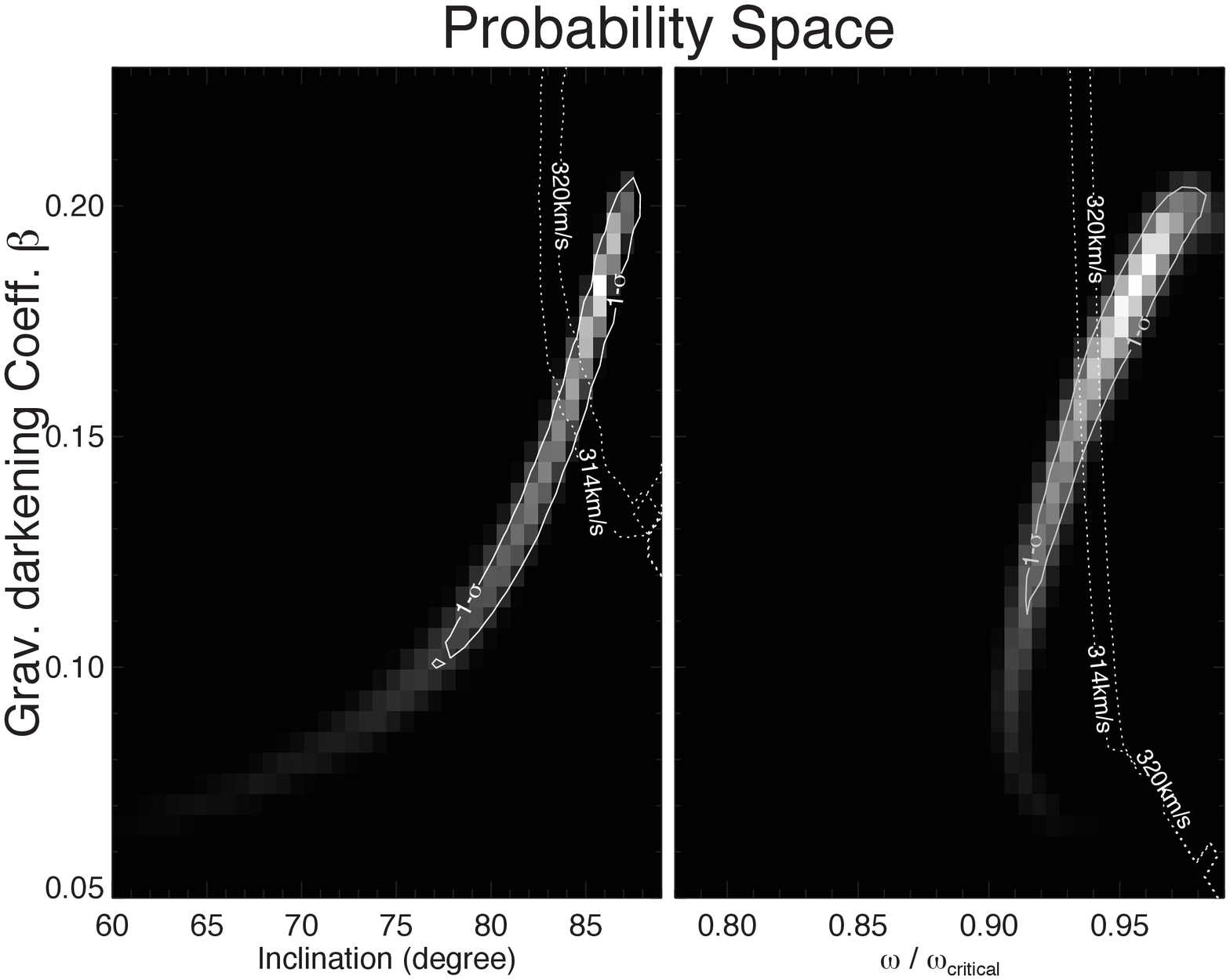}
}
\hphantom{.....}
\caption{ 
Probability spaces of $\alpha$ Leo show the degeneracy between stellar parameters. All the notations are the same as in the probability spaces of \betcas (see Fig. 2). The strong elongation of the contours in both panels suggest strong correlation between these parameters. The dashed lines show the \vsini range from McAlister et al. (2005), which intersects the probability contours with smaller common areas.
 \label{alfleo1}}
\end{center}
\end{figure}
\begin{deluxetable}{lcc}
\tabletypesize{\scriptsize}
\tablecaption{Best-fit and physical parameters of \alfleo}
\tablewidth{0pt}
\tablehead{ 
\colhead{Model Parameters} 
&\colhead{Modified von Zeipel model ($\beta$-free)}
&\colhead{von Zeipel model ($\beta$ = 0.25)}
 }
\startdata

Inclination (\degs)         			& $86.3^{+1.0}_{-1.6}$ 					& $87.5^{+0.2}_{-0.1}$				 \\
Position Angle (\degs) 			&  $258^{+2}_{-1}$						& $259^{+1}_{-2}$				 \\
\tpol (\K)                				&  $14520^{+550}_{-690}$ 				& $16190^{+150}_{-110}$ 			\\
\rpol (\mas)          				&  $0.617^{+0.010}_{-0.009}$          			&$0.605^{+0.001}_{-0.001}$			\\
\wratio    						&  $0.962^{+0.014}_{-0.026}$ 				& $0.969^{+0.001}_{-0.002}$	 		\\
$\beta$                          			& $0.188^{+0.012}_{-0.029}$	 			& 0.25 (fixed)\\

\hline
\hline
Derived Physical Parameters		&							&						\\
\hline
\teq  (\K)                				&  $11010^{+420}_{-520}$ 				& $10920^{+100}_{-70}$ 			\\
\req  (\rsun)    					&  $4.21^{+0.07}_{-0.06}$ 				& $4.17^{+0.007}_{-0.006}$ 			\\
\rpol (\rsun)   					&  $3.22^{+0.05}_{-0.04}$					& $3.16^{+0.005}_{-0.004}$			 \\
Bolometric luminosity \lbol (\lsun) 	& $341^{+27}_{-28}$ 					& $431^{+18}_{-9}$				 \\
Apparent effective temperature \tapp(\K)	& 12080 							&12650					\\
Apparent luminosity \lapp(\lsun)  	& 252	 							& 294						\\
Model \vsini (\kms )\tablenotemark{a}	& $336^{+16}_{-24}$ 					& $346^{+1}_{-2}$		 		\\
Rotation rate (rot/day)			&$1.64^{+0.02}_{-0.04}$					&$1.70^{+0.01}_{-0.01}$			\\
Model mass (\msun)\tablenotemark{b}	&$4.15 \pm 0.06$						&$4.52 \pm 0.05$			\\ 
Oblateness mass (\msun)	\tablenotemark{c}	&$3.66^{+0.79}_{-0.28} $					&$3.44^{+0.08}_{-0.01}$		\\
Age (\Gyr)\tablenotemark{b}		&$0.09 \pm 0.02$						&$0.05 \pm 0.01$			\\
Model $\emph{V}$ Magnitude\tablenotemark{d}    & $1.393^{+0.002}_{-0.005}$ 	& $1.329^{+0.017}_{-0.021}$					\\
Model $\emph{H}$ Magnitude\tablenotemark{d}    & $1.578^{+0.004}_{-0.006}$ 	& $1.550^{+0.012}_{-0.015}$					\\

\hline
\hline
$\chi^{2}$ of various data			&							&						\\
\hline
Total $\chi^2_{\nu}$        			&1.32  	 					& 2.57					 \\
Vis$^2$ $\chi^2_{\nu}$ 			&0.76   						& 1.26 					 \\
CP $\chi^2_{\nu}$          			&1.97    						& 3.80					\\
T3amp $\chi^2_{\nu}$   			&0.92    						& 1.52					\\

\hline
\hline
Physical Parameters from the literature	&							&						\\
\hline
[Fe/H]\tablenotemark{e}  & \multicolumn{2}{c}{0.0} \\
Distance (\pc)\tablenotemark{f} & \multicolumn{2}{c}{24.31} \\

\enddata
\tablenotetext{a}{Observed \vsini = 317 $\pm$ 3\kms \citep{mca05}}
\tablenotetext{b}{Based on the $Y^2$ stellar evolution model \citep{yi01, yi03, dem04}.}
\tablenotetext{c}{\cite{zha09}}
\tablenotetext{d}{\emph{V}mag = 1.391 $\pm$ 0.007 \citep{kha09}, 
\emph{H}mag = 1.658 $\pm$ 0.186 \citep{cut03}, 1.57 $\pm$ 0.02 \citep{duc02} }
\tablenotetext{e}{\cite{gra01} }
\tablenotetext{f}{\cite{van07} }
\label{alfleo_tab}
\end{deluxetable}

\section{Imaging Of Rapid Rotators}
Interferometric data contains information of the Fourier Transform of
the projected surface brightness of sources. Therefore, in theory, a
stellar image can be reconstructed directly from the data. But in
reality because of the finiteness of UV coverage and uncertainty from
measurements, many different images fit well to the same
interferometric data.  We use the application
'Markov-Chain Imager for Optical Interferometry' (MACIM; \cite{ire06})
to construct images for \betcas and \alfleo. It is usually difficult
to image nearly point-symmetric objects because the closure phases
will be close to either 0 or 180 degrees, making it harder to
constrain the detailed structure. \betcas is close to pole-on and
\alfleo is almost equator-on, which are two cases of the
point-symmetry.

One strategy to image these kinds of stars is to take advantage of
some prior knowledge. Stars are confined in certain area with
elliptical shapes approximately. Therefore we employ a prior image
which is an ellipse with uniform surface brightness. The spatial and
geometric parameters of the ellipse come from the model fitting.
The detailed process can be found in \cite{mon07}.

The left panel of Figure 4 shows the reconstructed image of
\betcas. The reduced $\chi^2$ of the image is 1.20, comparable to our best-fit models. 
We overplot
longitudes and latitudes with solid lines from the model and include contours
of surface brightness temperatures with dashed lines. The right panel
shows the image from the model fitting, overplotted with the surface
brightness temperature contours from the model. Because of the
inclination angle, the surface brightness temperature contours do not
coincide with latitude contours. We find that the two images are
consistent with each other in general. The images show a center bright
region which is one pole of \betcas. The surface brightness drops
gradually towards the edge due to gravity darkening. One may also
notice limb-darkening at the edge of the stellar image.

\begin{figure}[thb]
\begin{center}
{
\includegraphics[width=7.5in] {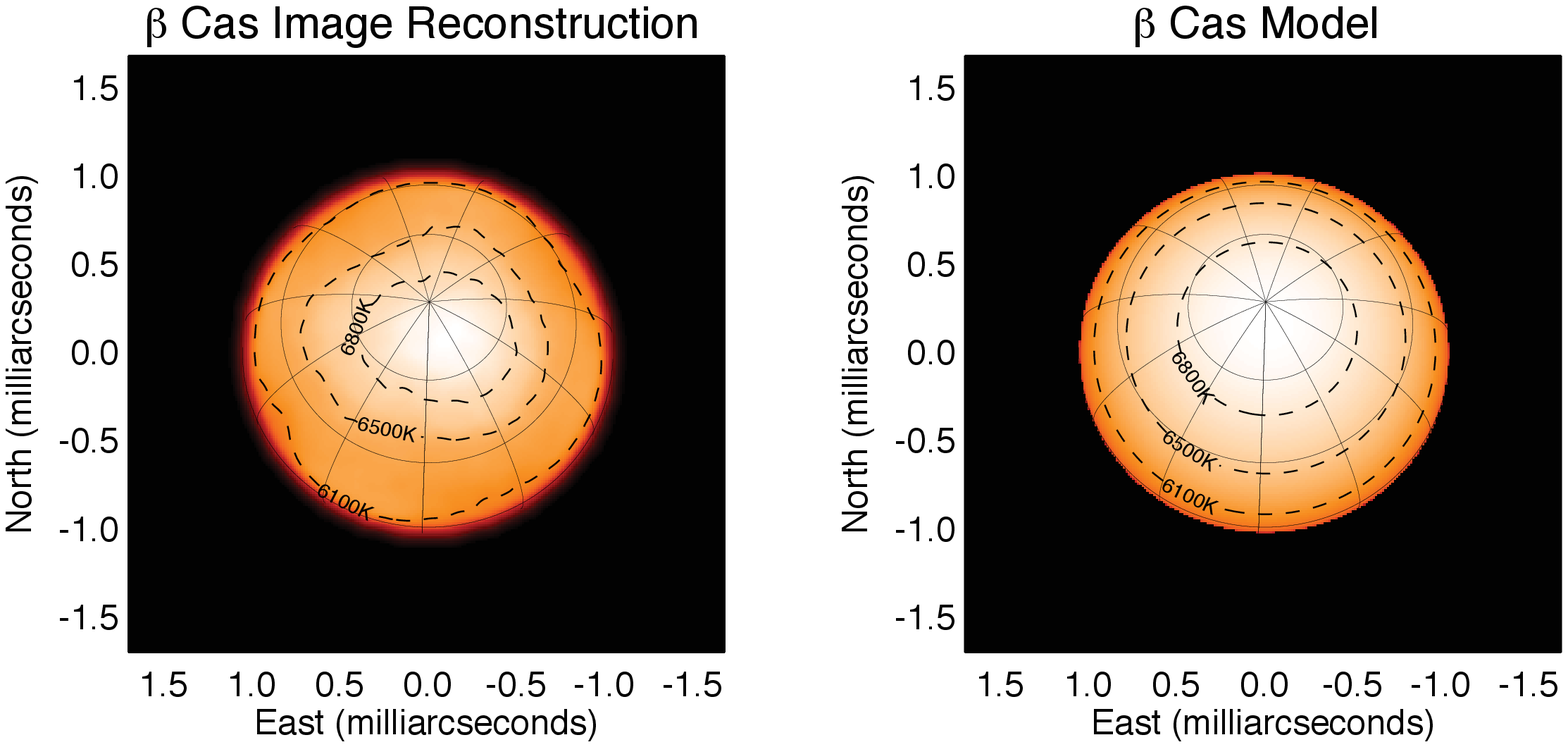}
}
\hphantom{.....}
\caption{ 
Images of \betcas. The left one shows the surface intensity distribution of \betcas from MACIM, overplotted with latitudes and longitudes from the model. The angular resolution is 0.57 \mas (milli-arcsecond). The dashed contours represent the surface brightness temperatures of the image. The right one shows the image from model fitting, overplotted with brightness temperature contour from the model. The reduced $\chi^2$ of the images from MACIM and model fitting are 1.20 and 1.36. 
 \label{betacas2}}
\end{center}
\end{figure}

The left panel of Figure 5 shows the image of \alfleo with latitudes
and longitudes from the model, and surface brightness temperature
contours. The reduced $\chi^2$ of the image is 0.78. The right one
shows the image from model fitting.  As
opposed to \betcas, \alfleo is almost equator-on and the dark equator
stretches along the North-South direction. One noticeable phenomenon
is that the poles are not located exactly in the hot region. This is
because in this particular case the poles at the stellar image edge
look cooler due to lime-darkening, causing the brightest regions to
shift towards the center of the image.

\begin{figure}[thb]
\begin{center}
{
\includegraphics[width=7.5in] {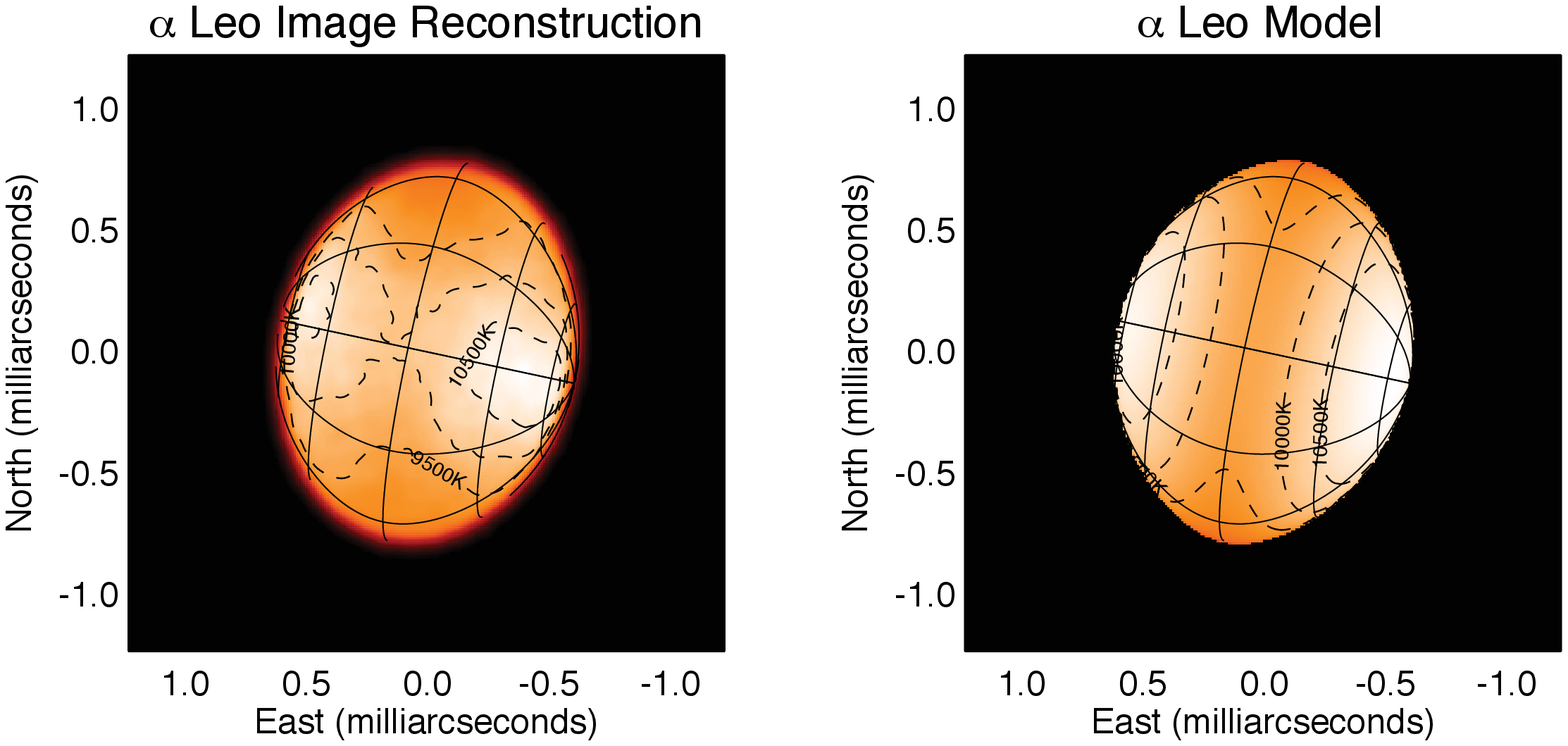}
}
\hphantom{.....}
\caption{ 
Images of \alfleo. The notations are all the same as those in images of \betcas (see Fig. 4). The angular resolution is 0.55 \mas. The reduced $\chi^2$ of the images from MACIM and model fitting are 0.78 and 1.32. 
 \label{betacas3}}
\end{center}
\end{figure}

\section{Stellar Evolution Tracks}
One interesting topic for rapidly rotating stars is to locate their
positions on the Hertzsprung\--Russell (HR) diagram and compare with
stellar models. This topic contains two issues. First, traditional
photometry observations only see the apparent luminosities \lapp and
apparent effective temperatures \tapp which depend on stellar
inclination angles; the bolometric luminosities \lbol of rapid
rotators are hidden from the observers. Interferometric observations
allow us to construct 2-D surface models of stars, thus to obtain the
\lbol \citep{zha09}. We obtain the gravity and temperature
distributions across the stellar surface from the model fitting. From
Kurucz model, we are able to retrieve intensities from each patch of
stellar surface, and then integrate the radiation all over the star to
obtain the bolometric luminosity \lbol\footnote{The "overall effective
temperature" \tbol can be estimated from the \lbol divided by the
total surface area; However, in the case of a rapid rotator, this overall
effective temperature is just a definition with limited physical meaning,
so it is not used to infer the masses or ages of stars in this
paper.}. By comparison we also compute an inclination curve which
shows stellar \lapp and \tapp as a function of the inclination angle, and
we can mark the one corresponding to its inclination from the model
fitting. The \lapp can be calculated by \lapp = 4$\pi d^2$\fbol, where
d is the distance and \fbol \ is the bolometric flux computed by
integrating flux from each grid over the projected area. Then the
\tapp is obtained by $\sigma$(\tapp)$^4$ = $\pi d^2$\fbol /\aproj,
where \aproj \ is the projected area.

Second, typical HR diagrams are constructed for non-rotating stars, it
is inappropriate to place a rapid rotating star on such diagrams. A
rapidly rotating star shows a little lower \lbol than \lnr from its
non-rotating equivalent (an imaginary spherical star which a rapid
rotator would turn out to be if it spins down to no angular velocity),
meaning a rotating star will evolve as a lower mass star on HR
diagram. Therefore the interpreted mass and age from the rotating star
deviates from the true values. To partially solve this problem, one has to
convert the properties of a  rapidly rotating star to its non-rotating
equivalent. Studies have shown that the bolometric luminosity and
polar radius do not change much as a star spins up. Following this,
we alter the traditional HR diagram
to a new one with axes of bolometric luminosity and polar radius (\lr
diagram), and locate rotating stars on the new diagram to infer the
mass and age \citep[][private communication, 2010]{pet06}. To compare with the astronomy-friendly HR diagrams, one
can also translate these two values of non-rotating equivalents into
\lnr and \tnr.

The left panels of Fig. 6 and 7 show \betcas and \alfleo on \lr
diagrams from $Y^2$ model \citep{yi01, yi03, dem04}. The cross and
square symbols represent the bolometric luminosity and polar radius
before and after the rotational correction respectively
\citep{sac70}. The corrections are trivial: \lnr and \rpolnr decrease
by 5.5\% and 1.3\% respectively for a 2 solar mass star as it spins up
to close to critical speed. So on \lr diagrams one may even directly
use \lbol and \rpol of a rotating star for rough interpretations of
its mass and age.  We have begun work on a more exact
formulation using a new grid of rotating models, but this is the subject of 
future detailed paper.

The traditional HR diagrams are shown in the right panels of Fig. 6
and 7. The solid lines are the inclination curves, which show the
\lapp and \tapp as a function of inclination angles. The star symbols
on the curve represent the estimated inclination angles. The square
symbols stand for \lnr and \tnr of the non-rotating equivalent. The
position of non-rotating equivalent on HR diagram deviates severely
from the position of the rapidly rotating equivalent based on its
apparent values. For instance, Regulus would be about 0.08 Gyr older
and 0.5 \msun less massive from its \lapp and \tapp than from \lnr and
\tnr. So we strongly recommend to correct for the effects of rotation
when placing a rapidly rotating star on HR diagram.  \cite{zha09}
didn't adopt this correction, which may lead to an additional error in
determining age and mass of rapidly rotating stars.

\begin{figure}[thb]
\begin{center}
{
\includegraphics[width=3.2in,trim=34mm 22mm 25mm 25mm,clip=true] {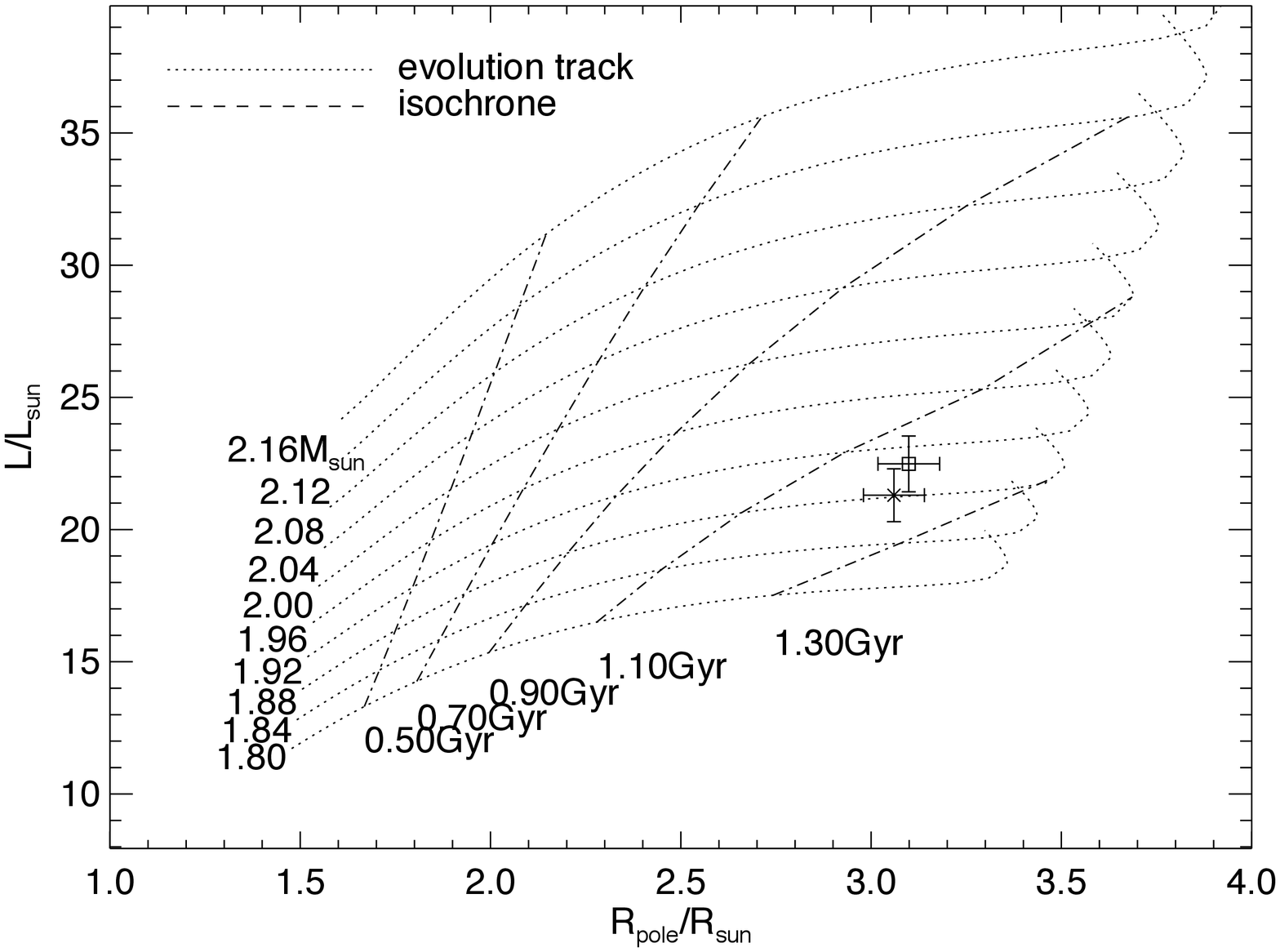}
\includegraphics[width=3.2in,trim=34mm 22mm 25mm 25mm,clip=true] {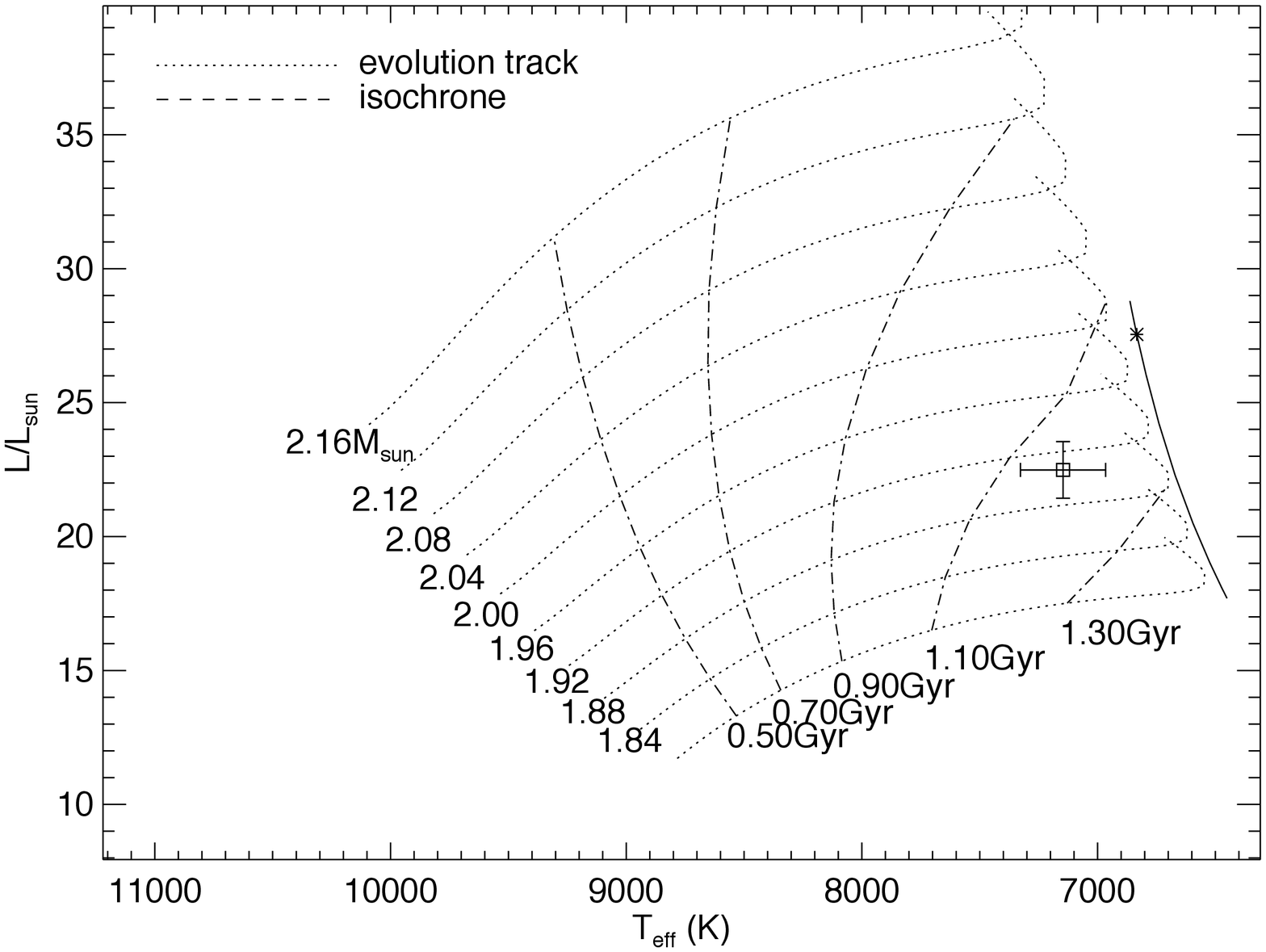}
}
\hphantom{.....}
\caption{ 
\betcas positions on \lr (left) and Hertzsprung\--Russell (right) diagrams based on $Y^2$ model \citep{yi01, yi03, dem04}. In the left panel, the cross symbol with error bar stands for the rapidly rotating \betcas based on its \lbol and polar radius from modified von Zeipel model fitting. The square symbol with error bar is the non-rotating equivalent of \betcas, the corrections of \lbol and polar radius because of rotation is adopted from \cite{sac70}. In the right panel, the solid line is the inclination curve, which shows how \lapp and \tapp change as a function of inclination angles. The star symbol is \betcas with its estimated inclination angle. The meaning of the square symbol is the same as in the left panel (see Section 5). 
 \label{betacas4}}
\end{center}
\end{figure}

\begin{figure}[thb]
\begin{center}
{
\includegraphics[width=3.2in,trim=29mm 22mm 25mm 25mm,clip=true] {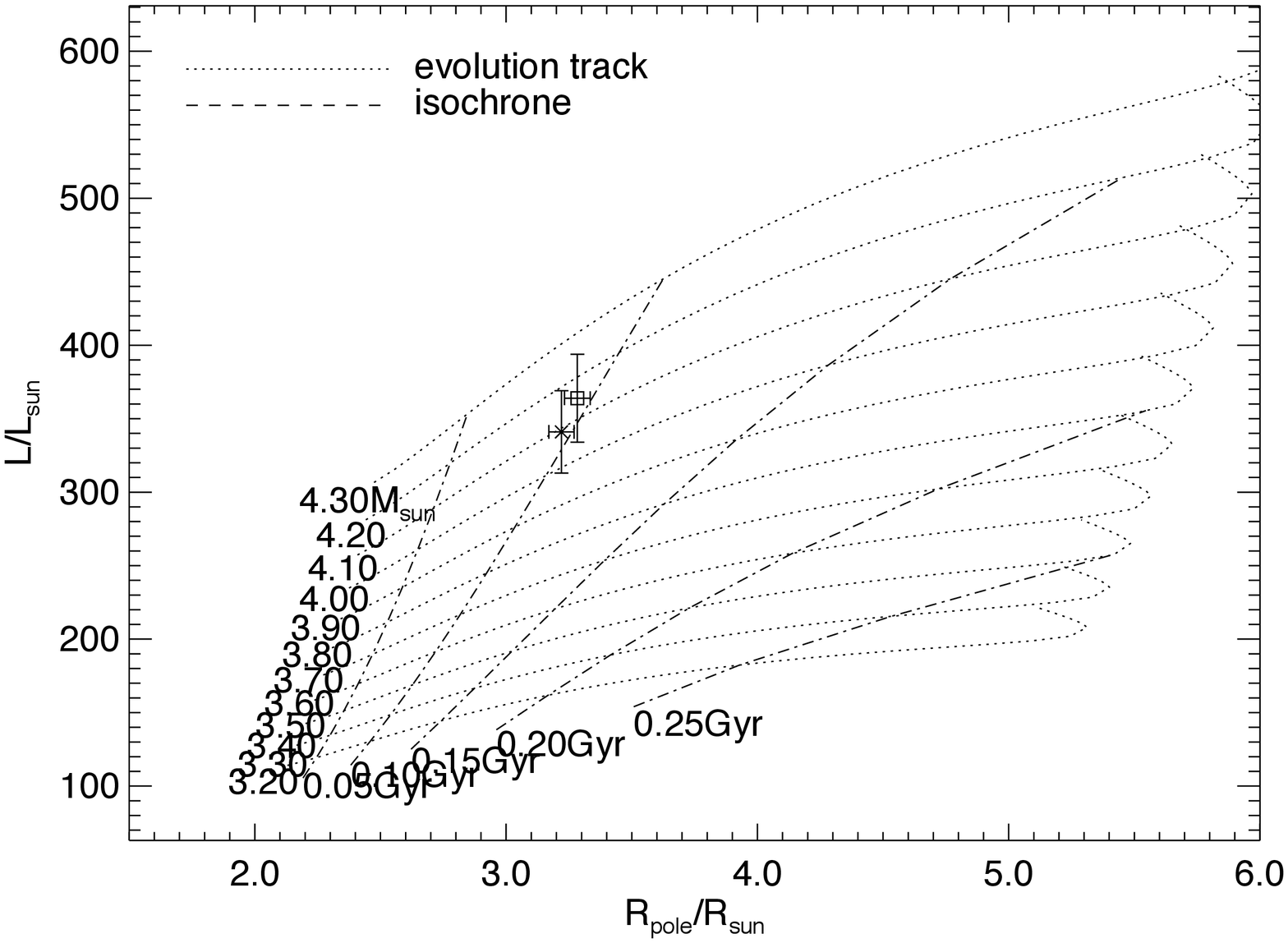}
\includegraphics[width=3.2in,trim=29mm 22mm 25mm 25mm,clip=true] {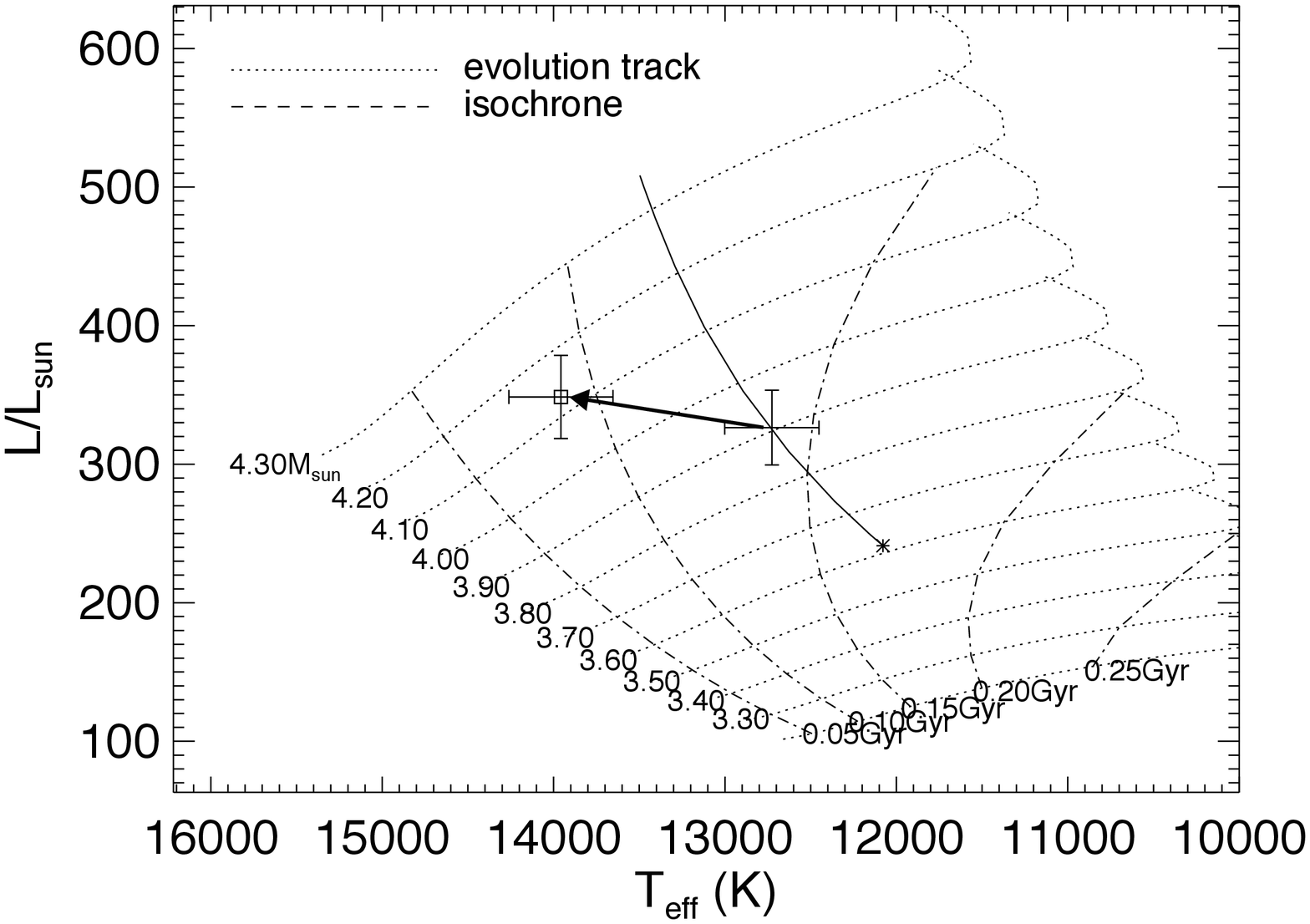}
}
\hphantom{.....}
\caption{ 
\alfleo position on \lr and HR diagrams based on $Y^2$ model. The notations are the same as those on diagrams of \betcas (see Fig. 6).
 \label{alfleo2}}
\end{center}
\end{figure}

\section{Discussion}
\subsection{Stellar Core-Envelope Coupling}

Measuring \wratio \ as a function of age provides a way of studying
the coupling between the stellar core and envelope in terms of angular
momentum. As a star evolves along the main sequence, the core
contracts and spins up due to the conservation of the angular
momentum, while the spherical-shell envelope expands and spins
down. \wcrit \ also drops as the star expands. Given the initial
rotational conditions and the evolution of stellar inner structure,
the evolution of \wratio \ depends only on how much the core and
envelope are coupled. In the case when the core and envelope are not
coupled, the angular velocity of the envelope changes roughly
proportional to $\rm{R}^{-2}$. The ratio \wcrit \ is proportional to
$\rm{R}^{-1.5}$. So \wratio \ decreases roughly as $\rm{R}^{-0.5}$ as
a star expands. While in the other extreme case of solid body
rotation, namely the core and envelope are fully coupled, the core
transfers the most angular momentum to the envelope, and \wratio \ may
increase as a star expands. We can also predict its value in the past,
knowing the current \wratio.

One critical component in the discussion above is the evolution model
of stellar inner structure. While several such models are available
for non-rotating stars, we can not find one for the general case of
rotating stars. We justify that a non-rotating stellar model is a good
approximation for calculating evolution of internal density profiles because
rotation has very little effect on iso-potential surfaces inside the
star.  For instance, a rapidly rotating star with \wratio \ = 0.9, its
equatorial radius is elongated by only 21.6\%, but gravity quickly
dominates as one looks deep into the star. This means \wcrit \ is much
larger than angular velocity at certain radius and smaller, and the
structure can again be approximately described by a non-rotating
stellar model. So in the following calculation we adopt a non-rotating
stellar model\footnote{EZ-Web
http://www.astro.wisc.edu/$\sim$townsend/static.php?ref=ez-web is a
web-browser interface to the EZ evolution code \citep{pax04},
developed and maintained by Rich Townsend.}

By computing how the moment of inertia changes with time, we are able
to calculate the evolution of \wratio for a 1.9$\msun$ non-rotating star (Fig. 8).
In the left panel, all the
values are normalized to their initial values. The solid line shows
the evolution of the stellar radius, the dotted and dashed lines show
the evolution of the ratio \wratio \ when the core and envelope are
fully coupled and uncoupled. When the core and envelope are uncoupled,
the ratio drops as the star expands as expected. When the core and
envelope are fully coupled, the ratio actually increases a little due
to the transference of angular momentum from the core to the
envelope. This result may explain high \wratio \ value of  \betcas.

In the right panel, we use the ratio \wratio \ = 0.92 from model
fitting as the current value of \betcas, and trace back to its
previous values in the extreme cases of full-coupling and no
coupling. We notice that if the core and envelope are not well-coupled
(dashed line), the ratio will exceed the unit in the past, which is
not allowed. On the other hand if they are totally coupled (dotted
line), the ratio value remains below 1. Reading off the panel, \wratio
\ changes more rapidly in the past $\sim$ 0.5 \Gyr \ if the core and
envelope are not coupled. These results suggest that during the
stellar evolution of \betcas, the angular momentum is efficiently
transferred from the core to the envelope in the past 500 \rm Myr.
These results seem to confirm earlier findings by \citet{Danziger1972}
based on analysis of $v~sin~i$ statistics.

\begin{figure}[thb]
\begin{center}
{
\includegraphics[width=3.2in,trim=30mm 20mm 25mm 30mm,clip=true] {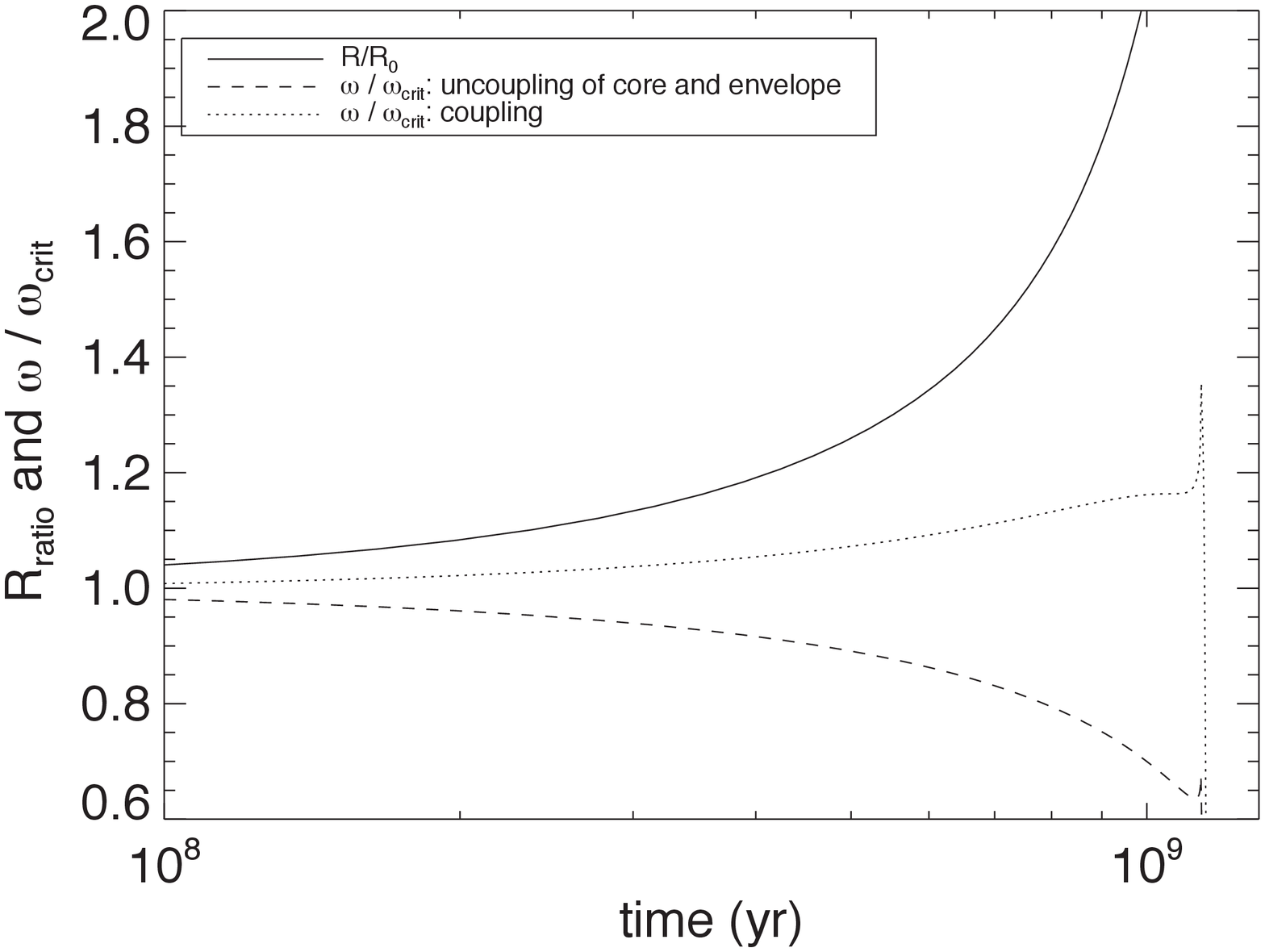}
\includegraphics[width=3.2in,trim=30mm 20mm 25mm 30mm,clip=true] {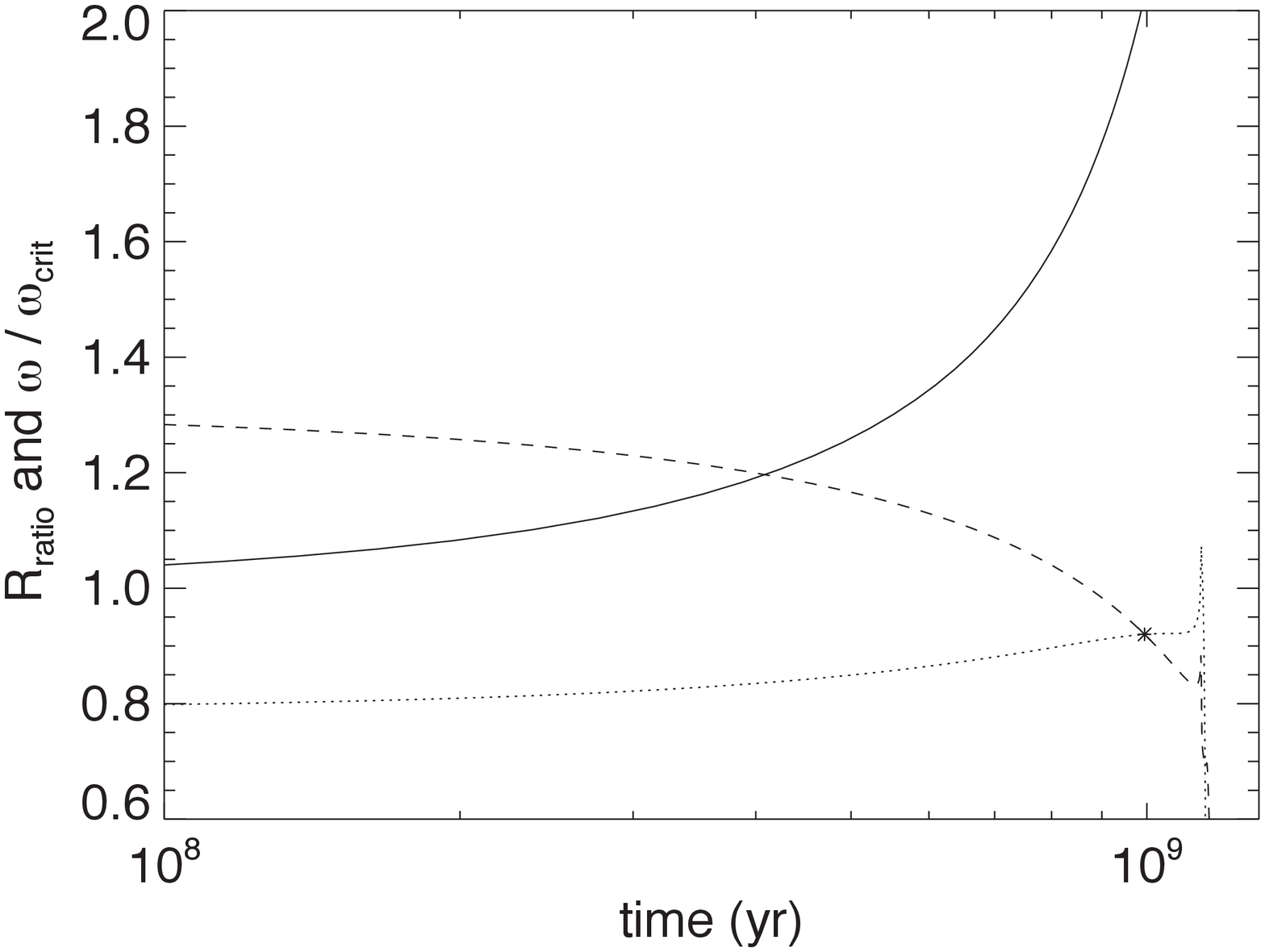}
}
\hphantom{.....}
\caption{ The evolution of stellar rotation. The model is adopted from
  the evolution of a 1.90 \msun \ non-rotating star 
  \citep[][the web-browser interface is developed and maintained by Rich Townsend]{pax04}. 
 The left
  panel: solid line is the ratio of the stellar radius to its value at
  the beginning of main sequence; dashed line is the ratio of \wratio
  \ ($\omega$ is angular velocity; \wcrit is the critical angular
  velocity when the centrifugal force balances the gravity at the
  equator) to its initial value when the core and envelope are not
  coupled; dotted line is the ratio when they are totally coupled. The
  right panel: using the current \wratio \ value (represented by
  asterisk) from model fitting, track back to its previous values
  assuming uncoupling and total coupling of the core and envelope.
 \label{rotation}}
\end{center}
\end{figure}

\subsection{Gravity Darkening Coefficient}

Von Zeipel brought up the idea of gravity darkening in 1924 and
predicted the standard value of $\beta$ to be 0.25 for stars with
fully radiative envelope. Our group have studied five rapid rotators
($\alpha$ Aql, $\alpha$ Cep, $\alpha$ Oph, $\alpha$ Leo, $\beta$ Cas)
up to now, four of them show non-standard Gravity darkening
coefficient ($\beta$) values from the modified von Zeipel model
fitting. $\alpha$ Oph was only fitted with $\beta$-fixed model because
of the high degeneracy between gravity darkening coefficient and
rotational speed due to its almost equator-on orientation
\citep{zha09}.

In Fig. 9 we plot the results of $\beta$ versus temperature for the
four targets with their gravity darkening coefficients obtained from
the modified von Zeipel model fitting. The shadow areas show the
temperature ranges from the pole to equator and the 1-$\sigma$ uncertainties of
$\beta$ from the model fitting for each star. For comparison, we also
plot the solid line representing the predicted relation between
$\beta$ and temperature adopted from \cite{cla00}. We digitize the
evolution plot of a 2 solar mass star in \cite{cla00} paper and extend
$\beta$ to high temperature 14500\K \ with $\beta$ fixed to 0.25.  
We should point out
that the predicted relation shifts a little to lower temperature for
stars with higher masses, but it is not a big issue in our case. For
$\alpha$ Cep, $\alpha$ Aql and $\beta$ Cas, their masses are close to
2 \msun, so they can share the same relation. $\alpha$ Leo is much
more massive than 2 \msun, the predicted curve shifts to low
temperature a little (less than 1000K).

\begin{figure}[thb]
\begin{center}
{
\includegraphics[width=7.5in,trim=27mm 20mm 0mm 20mm,clip=true] {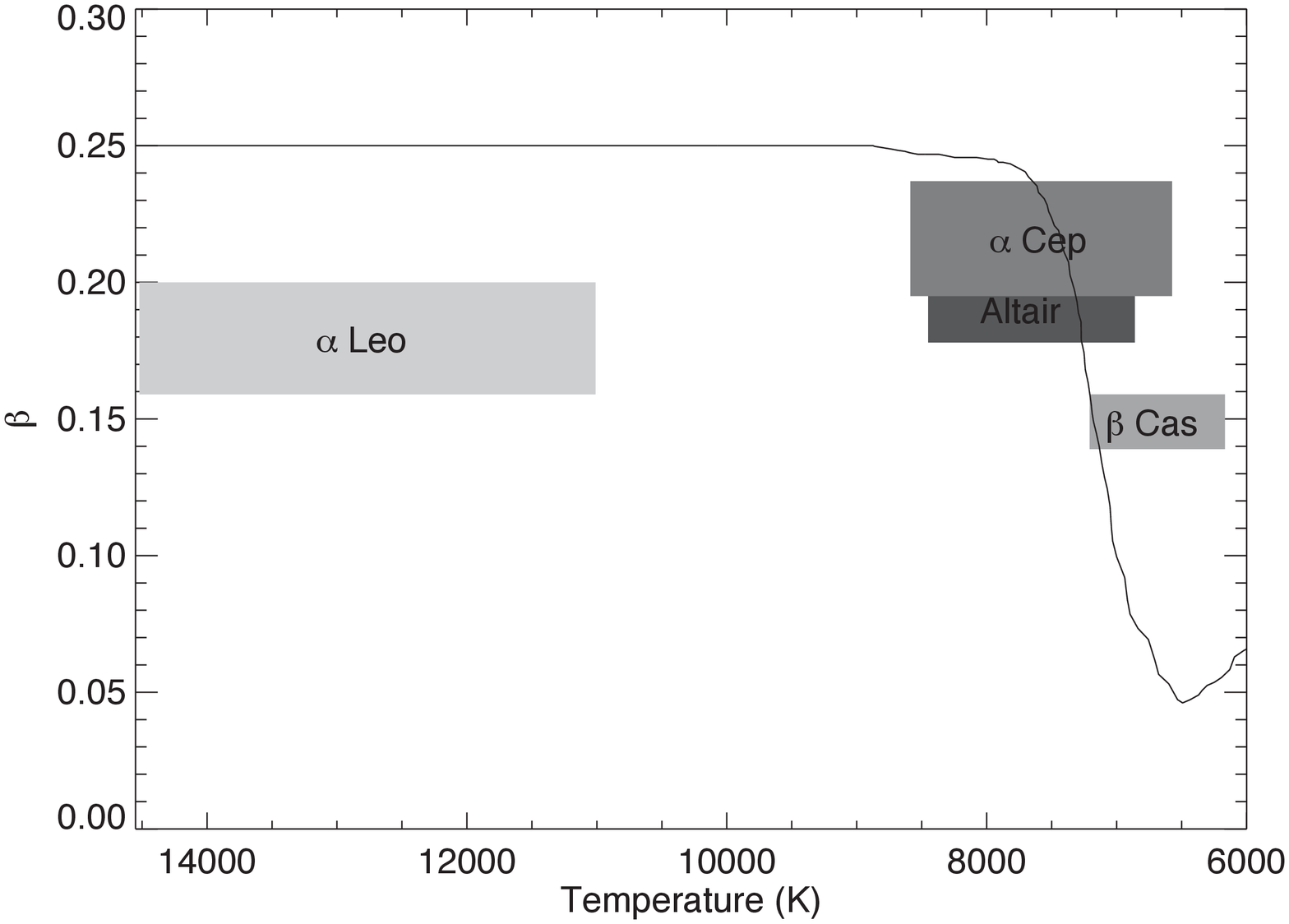}
}
\hphantom{.....}
\caption{ Gravity darkening coefficient ($\beta$) vs. temperature for four targets our group has studied. The solid line represents the theoretical relation between the gravity darkening coefficient $\beta$ and effective temperature, adopted from the evolution of a non-rotating 2 solar mass star \citep{cla00}. The curve is extended to higher temperature for comparison with \alfleo (see section 6.2 for details). The temperature range of each star contains temperature from the poles to equator. The $\beta$ range indicates the uncertainty from the model fitting.
 \label{claret}}
\end{center}
\end{figure}

Fig. 9 shows that $\alpha$ Cep, $\alpha$ Aql and $\beta$ Cas 
partially intercept the transition area of the predicted curve, meaning that the equatorial regions might
star to show convection. In our model fitting, we
use a single $\beta$ to describe the relation between the gravity and
temperature, instead of letting $\beta$ change as a function of
temperature. This may partially explain why these three stars have non-standard
$\beta$ values, because their poles could be radiation-dominated while
the equators convection-dominated, the resulting $\beta$ may be some
weighted values across the stellar surfaces. However the analysis here
is non-physical, a detailed stellar model that includes radiation and
convection in a rapidly rotating star is required to fully understand
the gravity darkening law of these stars with intermediate
temperatures.  

However $\alpha$ Leo has such high temperature range that even the equator
is supposed to be fully radiative theoretically. So the poles and
equator will share the same $\beta$ = 0.25, justifying the standard
von Zeipel model in this case. But our result still prefers
non-standard $\beta$ = $0.188^{+0.012}_{-0.029}$ . One possible
explanation is that even at such high temperature, the envelope is not
fully radiative. \cite{tas00} concludes that solid-body rotation is
impossible for a pseudo-barotrope in static radiative equilibrium. The
solid-body rotation will disrupt the constancy of the temperature and
pressure over the stellar surface, and cause the temperature and
pressure gradients between the equator and poles. The gradients will
induce a flow of matter which forms a permanent meridional circulation
and break down the strict radiative equilibrium. The matter flow may
further lead to the failure of our model assumption: solid-body
rotation. The material from higher latitudes carries less angular
momentum than those from lower latitudes. The meridional flows moving
towards higher or lower latitudes will speed up or slow down the
rotational speed of local material on their way, which triggers
differential rotation.

Another study from \cite{lov06} compares the effective temperature
distribution across the surface of a 6.5 \msun \ solid-body rotator
between a stellar evolution model with rotation (ROTORC) and von
Zeipel's law, and finds that the temperature distribution is shallower
in the model which is consistent with lower $\beta$ value we obtained
from \alfleo. A few observations on W UMa systems \citep{kit88, pan98}
roughly confirm von Zeipel's law, but with very large scatter. The
material flows on the surfaces of these stars are less complicated due
to an important feature of the binary systems: the stars are tidally
locked by their companions. Hence the stellar differential rotations
are effectively depressed and the resulting solid-body rotations are
well regulated. Therefore these stars may maintain radiation-dominated
envelopes which validate the standard von Zeipel model.

Based on the similar $\beta$ value found for all our objects and for $\alpha$~Leo in particular,
we recommend researchers adopt a new standard $\beta$=0.19 for future modeling of rapid rotating stars with radiative envelopes.

\section{Conclusion}

We have studied two rapid rotators with extreme spectral type: $\beta$
Cas and $\alpha$ Leo observed by CHARA-MIRC. By fitting the modified
von Zeipel model, namely the solid-body rotation model with
free-$\beta$ gravity darkening law, to observed infrared
interferometry data and \emph{V} and \emph{H} photometric fluxes, we
find both stars are rotating at close to critical speed: \wratio= 0.92
and 0.96. The fast rotations elongate their equators by 24\% and 30\%
compared with their poles, and their equatorial temperatures are 1000K
and 3000K cooler than their polar values. We estimated the mass of
\alfleo to be 4.15 $\pm$ 0.06 \msun \ from both \lr and HR diagrams
corrected for rotational effect, and it is much higher than 3.4 $\pm$
0.2 \msun found by \cite{mca05}. We have also reconstructed aperture
synthesis images using MACIM. The images are consistent with the
temperature distribution from the model fitting.

We discussed the evolution of \wratio. The ratio could increase or
decrease depending on how much stellar cores and envelopes are
coupled. In the case of fully coupling, \wratio \ increases a little
during main sequence and sub-giant branch due to the angular momentum transferred from the
core to the envelope. Our study on $\beta$ Cas, which is about 1.18
\Gyr old but still rotating at 92\% of its critical speed, suggests
the core and envelope are well coupled during the evolution.

All our targets from the modified von Zeipel model fitting prefer the
non-standard gravity darkening coefficients, especially in the case of
$\alpha$ Leo whose envelope should be fully radiative because of the
high surface temperature range 11010K - 14520K. One possible reason is
that solid-body rotation breaks down the constancy of temperature and
pressure on the stellar surface and induces meridional flow, which
violates strict radiative equilibrium. Furthermore the
meridional flow may result in differential rotation which causes the
failure of our solid-body rotation assumption. To explore this
possibility in the future, we will construct a differential rotation model to fit
observed high resolution spectra of these rapidly rotating stars. 
Until better models are created, we recommend using the empirically-determined gravity-darkening coefficient
$\beta$ = 0.19 for rapidly-rotating stars with radiative envelopes.

\acknowledgements We acknowledge interesting discussions with Antonio
Claret, Jason Aufdenberg, Chuck Cowley, and Chris Matzner when
preparing this manuscript. The CHARA Array is funded by the National
Science Foundation through NSF grants AST-0307562, AST-0606958,
AST-0908253 and by the Georgia State University. Funding for the MIRC
combiner came from the University of Michigan and observations were
supported through National Science Foundation grants AST-0352723,
AST-0707927, and AST-0807577.

\bibliography{apj_ref_mod}   
\bibliographystyle{apj}

\appendix
\addcontentsline{toc}{chapter}{Appendices}
\section{Appendix}

Visibility data and fitting results from one single night are shown in Fig. 10 and 11. The upper left panels show all 7 and 5 nights combined visibility data of \betcas and \alfleo respectively, overplotted with the visibility curves of uniform disks with diameters of major and minor axises from model fitting. The other panels show the model fitting and imaging results compared with one single night data. The date of that night of \betcas is Oct. 22nd 2009, and that of \alfleo is Dec 8th 2008.

\begin{figure}[thb]
\begin{center}
{
\includegraphics[width=2.7in,trim=20mm 20mm 20mm 20mm, clip=true] {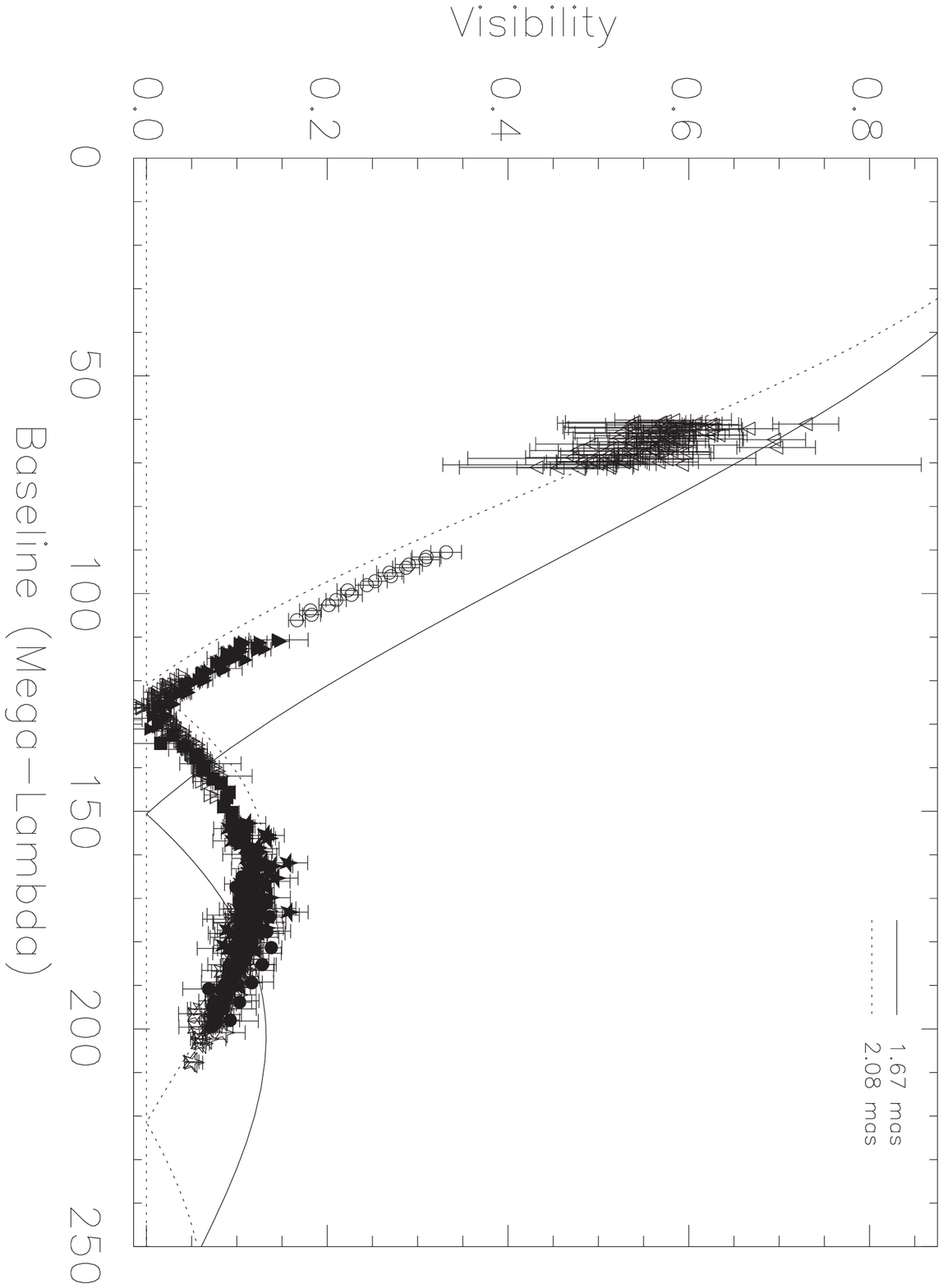}
\includegraphics[width=2.7in,trim=10mm 20mm 30mm 20mm, clip=true] {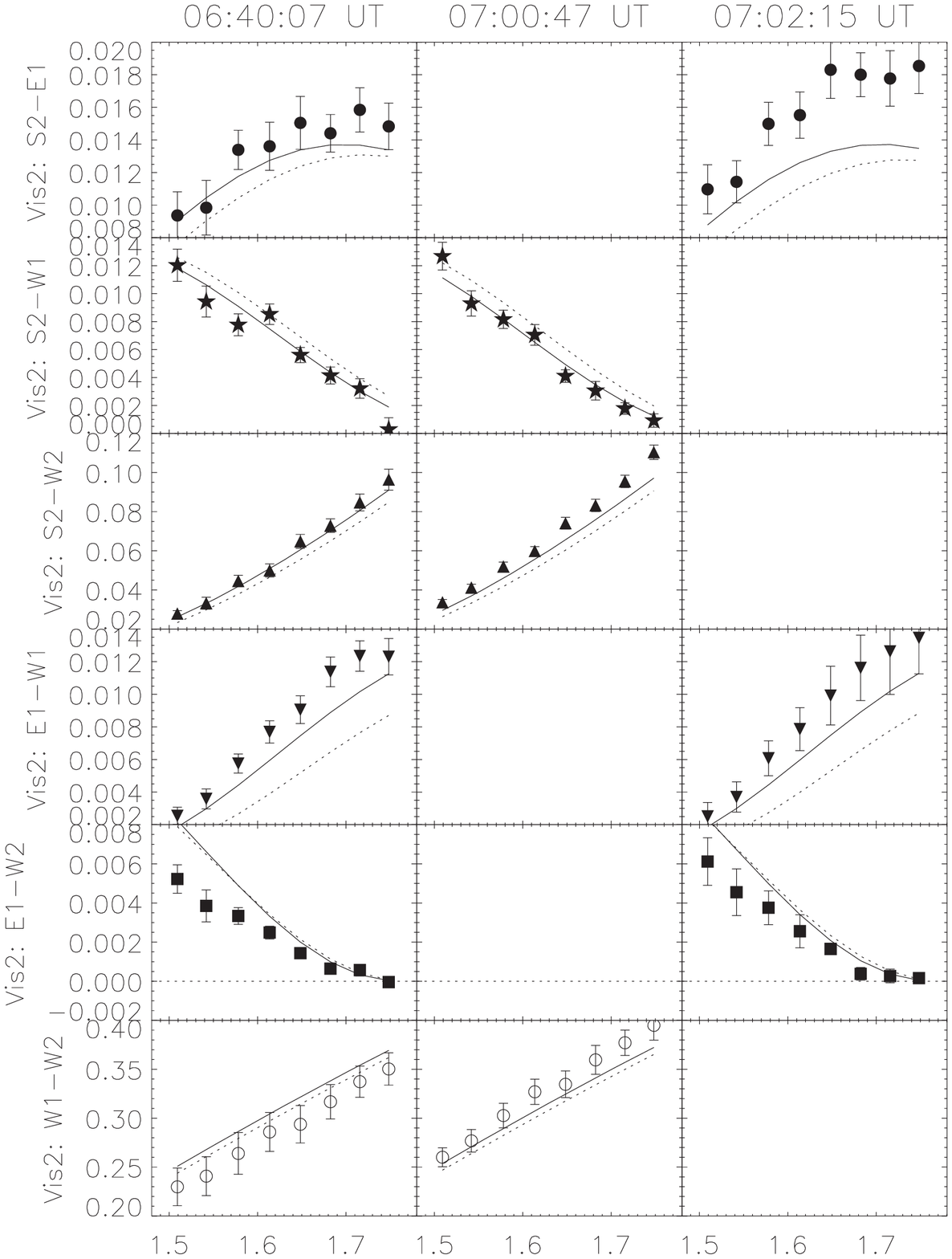}
\includegraphics[width=2.7in,trim=10mm 20mm 30mm 20mm, clip=true] {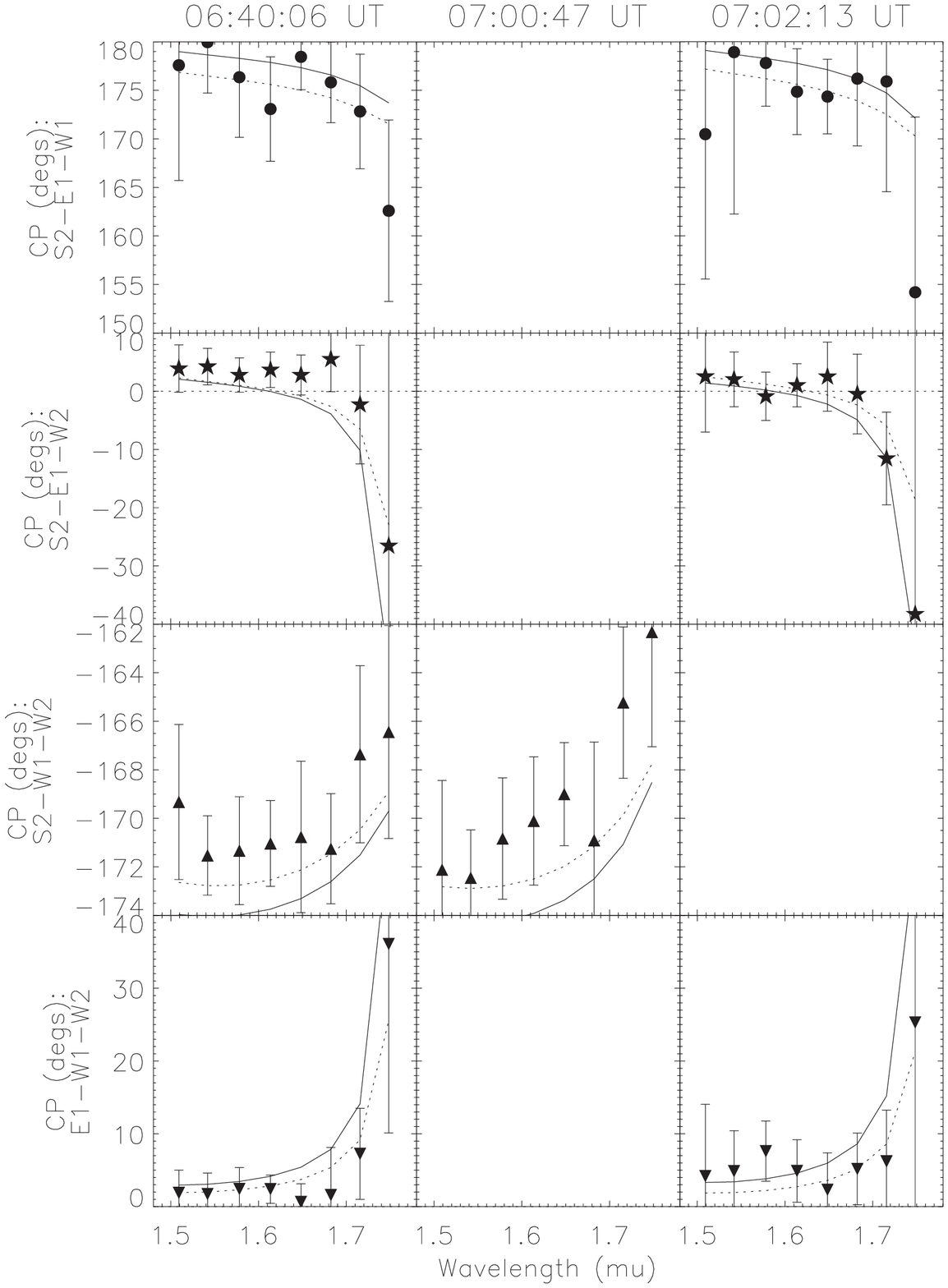}
\includegraphics[width=2.7in,trim=10mm 20mm 30mm 20mm, clip=true] {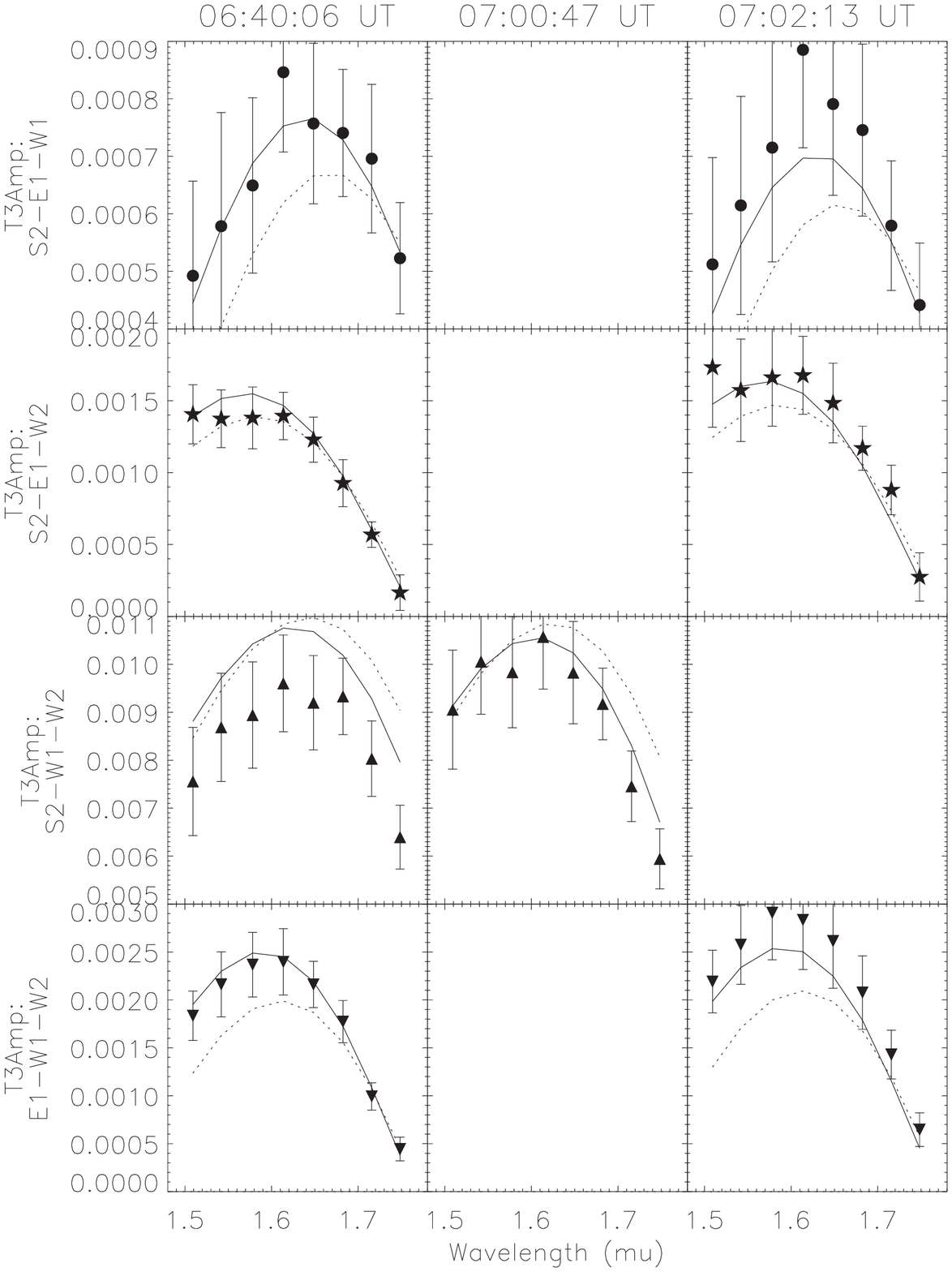}
}
\hphantom{.....}
\caption{Upper left panel: all seven nights visibility data of \betcas. The solid line and dotted line show the visibility curves of uniform disks with diameters of major and minor axis of \betcas from model fitting. The rest panels: the modified von Zeipel model (solid line) and MACIM image (dotted line, see section 4) vs. observed data (filled points with error bars) of \betcas from one single night. The reduced $\chi^2$ of model is 1.36 and that of image is 1.20. The eight data points in each sub-panel are from eight sub-channels of MIRC observation across $\emph{H}$ band. The x axis shows the wavelengths corresponding to the data points. The y axis shows which telescopes of CHARA have been used. 
 \label{betcas}}
\end{center}
\end{figure}

\begin{figure}[thb]
\begin{center}
{
\includegraphics[width=2.7in,trim=20mm 20mm 20mm 20mm, clip=true] {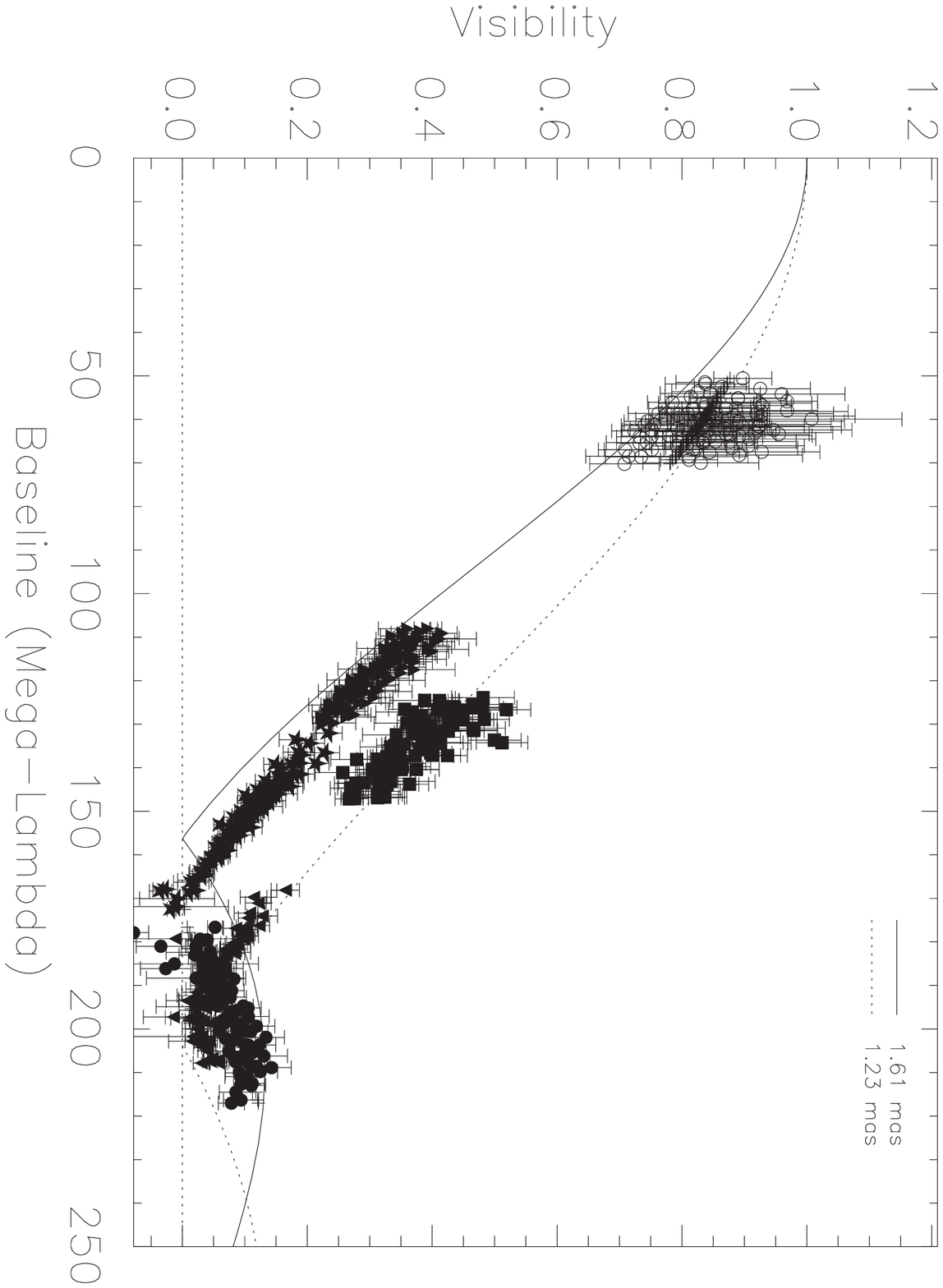}
\includegraphics[width=2.7in,trim=10mm 20mm 28mm 20mm, clip=true] {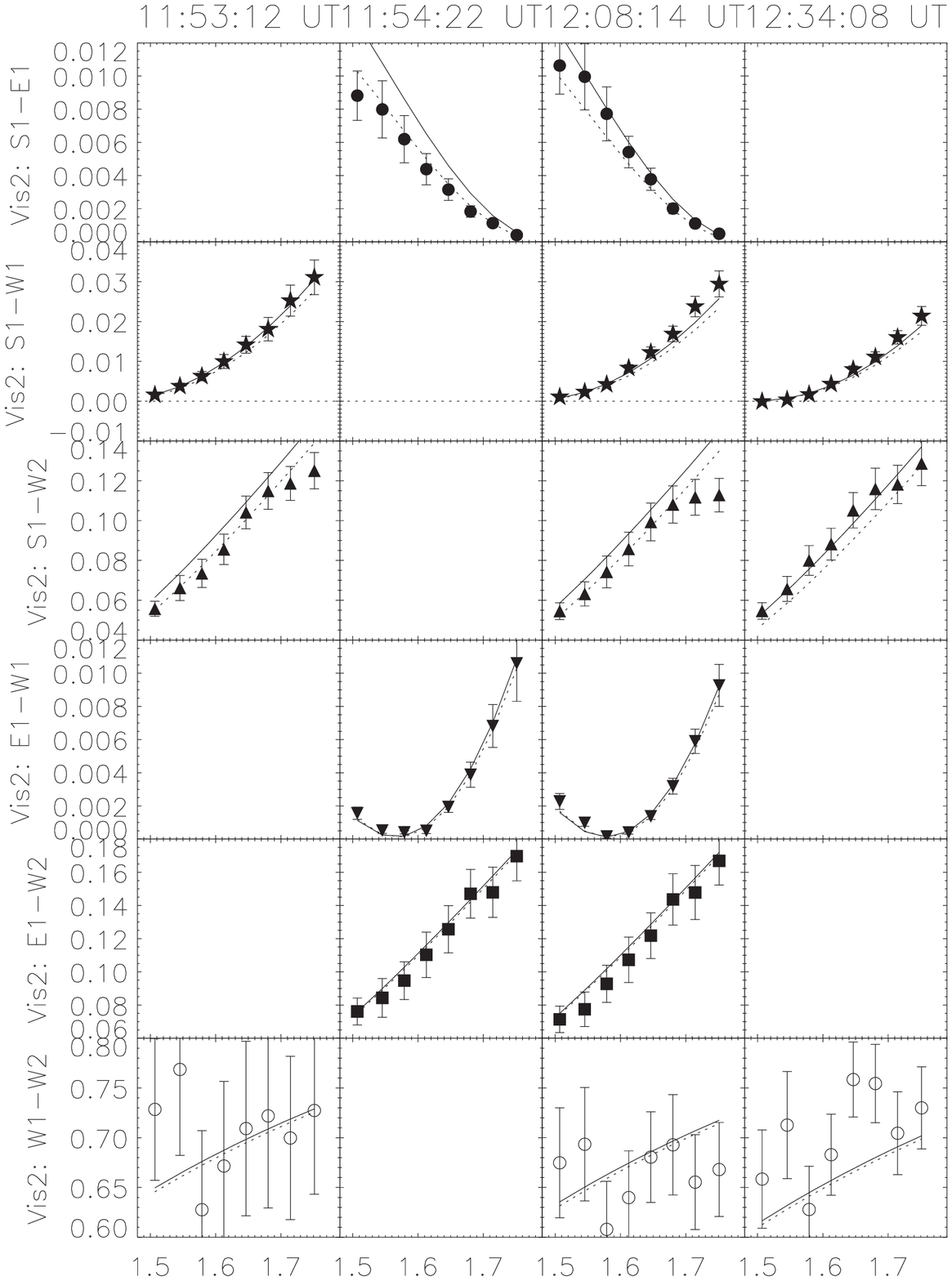}
\includegraphics[width=2.7in,trim=10mm 20mm 28mm 20mm, clip=true] {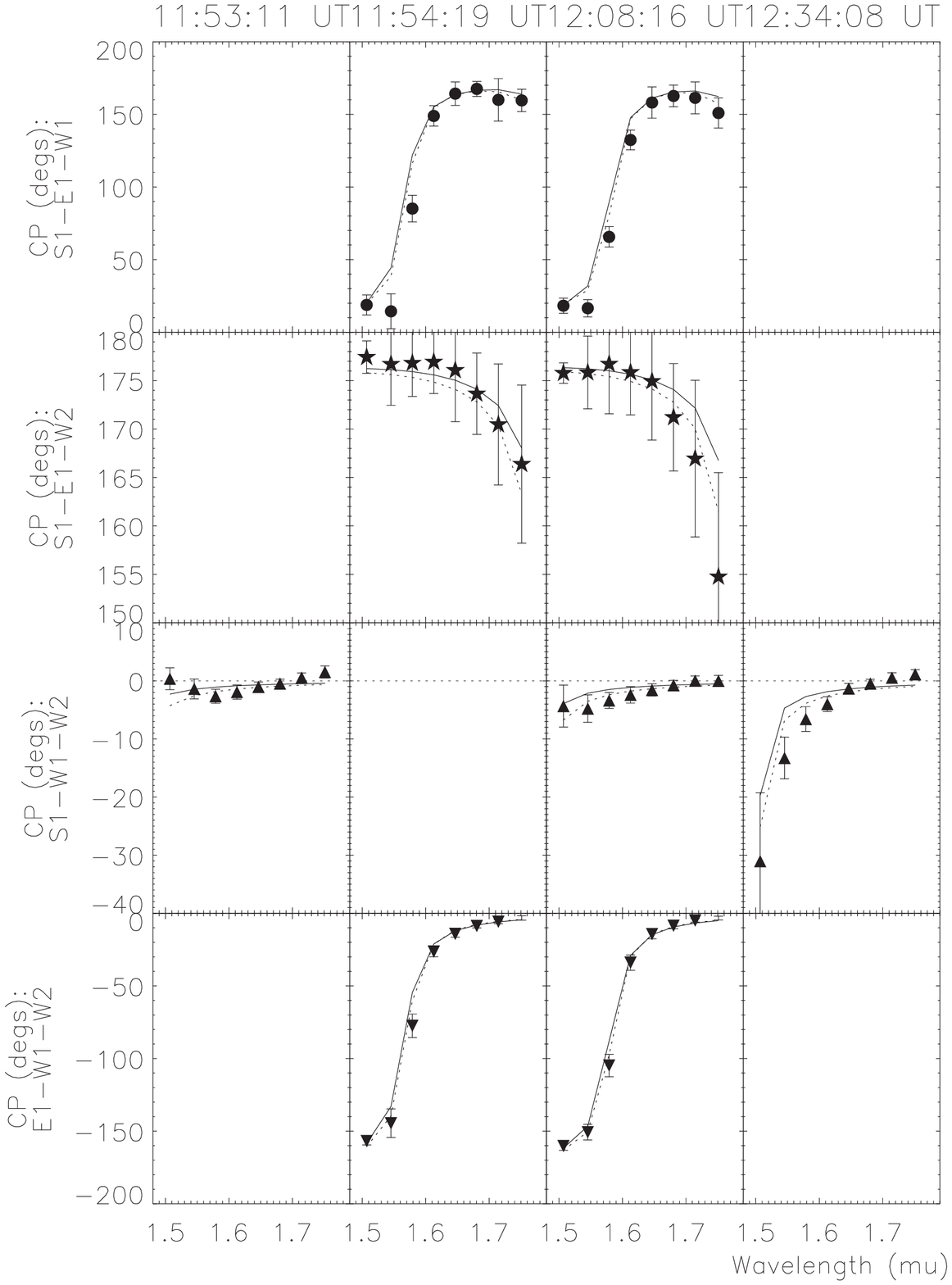}
\includegraphics[width=2.7in,trim=10mm 20mm 28mm 20mm, clip=true] {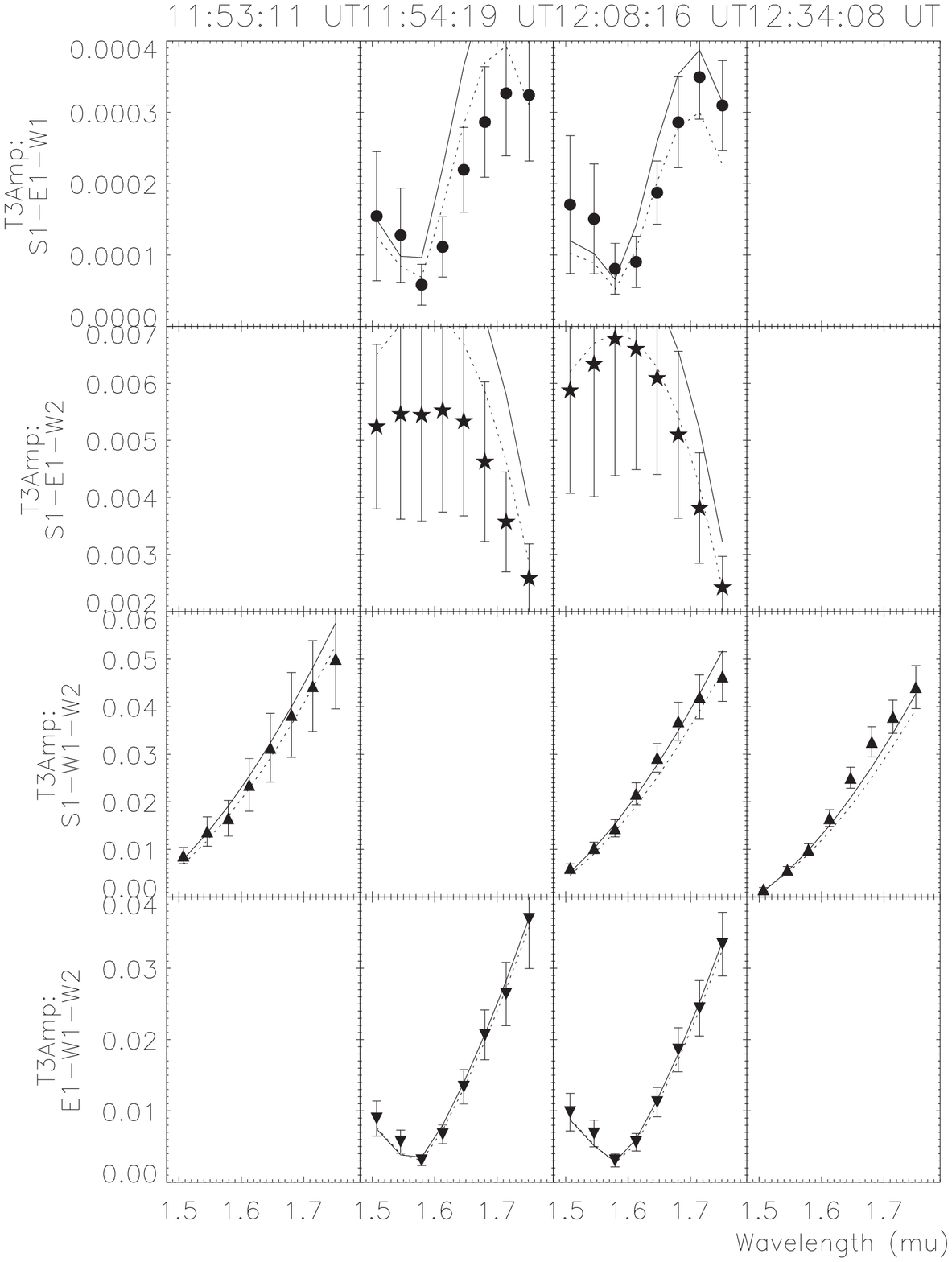}
}
\hphantom{.....}
\caption{The similar panels of \alfleo as those of \betcas in Fig. 10. The reduced $\chi^2$ of model is 1.32 and that of image is 0.78. 
 \label{alfleo}}
\end{center}
\end{figure}

\end{document}